\documentclass[11pt]{article}
\usepackage{float}
\usepackage[english]{babel}
\usepackage[utf8]{inputenc}
\usepackage{amsmath,amsthm,amssymb}
\usepackage{amsfonts}

\usepackage{bm}
\usepackage{subcaption}
\usepackage{comment}
\usepackage{booktabs}
\usepackage{algorithm}
\usepackage{algpseudocode}
\usepackage{dsfont}

\usepackage{multirow}
\usepackage{graphicx}
\usepackage{authblk}
\usepackage[sort,comma,authoryear,round]{natbib}
\usepackage[colorinlistoftodos]{todonotes}
\newcommand{\bigCI}{\mathrel{\text{\scalebox{1.07}{$\perp\mkern-10mu\perp$}}}}
\usepackage[bottom=0.8in,top=0.8in]{geometry}
\RequirePackage[colorlinks,citecolor=blue,urlcolor=blue, linkcolor = blue ]{hyperref}

\theoremstyle{definition}

\theoremstyle{plain}
\newtheorem{theorem}{Theorem}
\newtheorem{corollary}{Corollary}
\newtheorem{lemma}{Lemma}

\newtheorem{assumption}{Assumption}
\newtheorem{condition}{Condition}
\theoremstyle{remark}

\usepackage{tikz}

\usetikzlibrary{shapes,decorations,arrows,calc,arrows.meta,fit,positioning}
\usetikzlibrary{fit,calc,arrows.meta,patterns}

\tikzset{
    -Latex,auto,node distance =1 cm and 1 cm,semithick,
    state/.style ={ellipse, draw, minimum width = 0.7 cm},
    point/.style = {circle, draw, inner sep=0.04cm,fill,node contents={}},
    bidirected/.style={Latex-Latex,dashed},
    el/.style = {inner sep=2pt, align=left, sloped}
}
\tikzset{
treated/.style={circle, draw=blue, fill=blue, minimum size=6mm, inner sep=0pt},
control/.style={circle, draw=black, fill=white, minimum size=6mm, inner sep=0pt},
}

\def\beginmat{ \left( \begin{array} }
\def\endmat{ \end{array} \right) }

\newcommand{\bx}{\mathbf{x}}

\newcommand{\bX}{\mathbf{X}}
\newcommand{\bz}{\mathbf{z}}
\newcommand{\bZ}{\mathbf{Z}}

\newcommand{\bV}{\mathbf{V}}
\newcommand{\bA}{\mathbf{A}}

\newcommand{\bG}{\mathbf{G}}

\newcommand{\bY}{\mathbf{Y}}
\newcommand{\bW}{\mathbf{W}}

\newcommand{\bI}{\mathbf{I}}

\newcommand{\E}{\mathbb{E}}
\newcommand{\V}{\mathbb{V}}

\newcommand{\bepsilon}{\boldsymbol\epsilon}

\usepackage{bibunits}

\defaultbibliographystyle{Chicago}
\defaultbibliography{sample}

\title{Network-Constrained Matching and Randomization-Based Inference for Causal Effects}
\author[1]{Zhejia Dong\thanks{Email: zhejia\_dong@brown.edu}}
\author[1]{Youjin Lee\thanks{Email: youjin\_lee@brown.edu}}

\affil[1]{Department of Biostatistics, Brown University}
\date{}
\begin{document}
\begin{bibunit}
\maketitle

\begin{abstract}
Matching is widely used to mimic randomized experiments by forming matched sets in which treated and control units differ only randomly with respect to observed covariates. However, when the study population consists of interconnected units from a single network or a small number of networks, matching solely on observed covariates may produce matched units that are more closely connected in the network than would occur by chance. This increased network proximity within matched sets can induce spurious associations between treatment and outcome when both variables exhibit similar autocorrelation patterns on the network. 
To reduce spurious associations while preserving the validity of causal comparisons, we propose a new matching method that matches units with similar covariates subject to additional network-proximity constraints. For post-matching inference, we propose a randomization-based procedure for testing the sharp null hypothesis of no causal effect. The inference uses the asymptotic normal approximation and accommodates statistical dependence among test statistics obtained from each matched set without requiring explicit specifications of their correlation structures. 
We demonstrate the validity and utility of the proposed methods through simulation studies and apply them to the Framingham Heart Study. The matching method and subsequent inference procedure are implemented in the \texttt{R} package \texttt{netmatchRI}.
\end{abstract}

\section{Introduction}

\subsection{Spurious associations due to network dependence in causal inference}

Network dependence, also known as network autocorrelation, refers to statistical dependence within a variable due to network ties
~\citep{dow1982network,lee2021network}. In observational studies where human subjects are interconnected within a single network or a small number of networks, behaviors and health outcomes may propagate through network ties and introduce network dependence within each variable
~\citep{leenders2002modeling,ogburn2024causal}. When two or multiple variables exhibit similar network dependence, they may appear statistically correlated even in the absence of a true association. This phenomenon is known as \textit{spurious associations due to network dependence}~\citep{lee2021network}. 
Such spurious associations can lead to unique challenges for causal inference with network data, beyond those arising from (unmeasured) confounding. 

In observational network studies, both treatment assignments and outcomes may exhibit autocorrelation over the underlying network, often with similar dependence structures. For example, socially close friends may have similar baseline smoking behaviors and subsequent health outcomes (e.g., incidence of cardiovascular disease), both with similarity increasing with social proximity. In such settings, even when all confounding between smoking and health outcomes is adequately controlled, analyses based on socially proximate individuals may yield spurious causal conclusions between these two variables.  Distinct from confounding, network dependence does not arise from a common cause of these variables, and the resulting spurious associations generally cannot be eliminated by adjusting for network-related covariates or summary measures of network structure. One natural way to address both network dependence and confounding in causal inference is to consider how randomized experiments can be emulated using observational network data.

\subsection{Challenges of matching and post-matching inference under network dependence}

In observational studies, matching is widely used in causal inference to approximate randomized experiments by constructing matched sets of treated and control units with the same or similar covariates to reduce bias from observed confounders~\citep{stuart2010matching,rosenbaum2002observational}. 
However, in network data, units with similar covariates may also be proximate in the network space~\citep{shalizi2011homophily,khanam2023homophily}. Therefore, matching solely on covariates may produce matched units that are not randomly different with respect to their network distance, but instead are closer in the network. 
Such increased network closeness within matched sets may inadvertently exacerbate spurious associations. Matching design, therefore, should be modified to avoid matching units that are close in network space when their inclusion in the analysis may create spurious associations.

Statistical dependence induced by network ties may also affect post-matching inference. After causal comparison sets have been constructed through matching, randomization inference is often employed in the analysis stage because matching provides a design under which treatment assignments within each matched set may be analyzed as if randomized~\citep{pimentel2024covariate,zhu2023randomization,zhang2023randomization}. 
Such inference procedures are typically based on the assumptions that units within matched sets are independent, ensuring valid causal comparisons within matched sets, and that matched sets are independent of one another, enabling aggregation of evidence across sets using standard inferential methods. 
In observational network data, however, these two assumptions are generally not guaranteed. 
Consequently, standard randomization inference procedures that ignore potential dependence within and across matched sets may yield invalid conclusions.

\subsection{Our contribution}

In this work, we develop a matching design and randomization-based inference procedures for testing the sharp null hypothesis of no causal effect with observational network data. Our specific contributions are as follows.

First, we propose a new matching method that prevents network-proximate units from being matched into the same set while minimizing covariate dissimilarity. This design facilitates valid within-matched-set causal comparisons by reducing spurious associations due to network dependence. 
Second, we develop randomization-based inference procedures that account for dependence across matched sets without requiring explicit modeling of correlation structures.
We also provide several approaches for constructing covariance bounds that can be used for inference that relies on asymptotic normal approximation. 
The proposed matching methods and randomization-based inference procedures are implemented in the \texttt{R} package \texttt{netmatchRI}.
Overall, we provide a unified framework for matching-based randomization inference in observational network data where multiple variables may exhibit network dependence.

The remainder of this paper is organized as follows. In Section~\ref{sec:background}, we review the related literature and introduce the problem setup along with key assumptions. Section~\ref{sec:matching} proposes a new matching method.
Section~\ref{sec:inference} develops randomization-based inference procedures and the asymptotic normal approximation given the matched design. 
In Section~\ref{sec:simulation}, we present the finite-sample performance of the proposed methods through simulation studies under varying levels of network dependence. Section~\ref{sec:application} applies the methods to data from the Framingham Heart Study, a large longitudinal study that includes information on network ties among participants. Section~\ref{sec:discussion} concludes with a discussion of implications and future directions.

\section{Background and problem set-up }\label{sec:background}
\subsection{Literature review}

Despite the rapidly growing interest in causal inference with network data~\citep{tchetgen2012causal, aronow2017estimating, eckles2017design,   ogburn2024causal, basse2024randomization}, 
there remains a paucity of methodological literature on causal inference specifically addressing network dependence. 
Existing work largely falls into two categories. First, a substantial body of work studies treatment interference, in which one unit's treatment assignment affects another unit's potential outcome when the two units are connected through network ties~\citep{aronow2017estimating,forastiere2021identification,liu2014large,savje2021average}. 
Second, there is work on statistical dependence in network data, most of which focuses on dependence within a single variable, such as the outcome~\citep {basse2018model,leung2022causal} or the treatment assignment~\citep{toulis2025estimating}. These approaches often rely on the dependence structure within variables being known by design in randomized experiments~\citep{basse2018model} or being correctly modeled in observational studies~\citep{toulis2025estimating}. Our focus is on examining the presence of causal effects of treatments across the entire network, including treatment interference, while accounting for network dependence in multiple variables.

A closely related line of work in causal inference with dependent or autocorrelated data can be found in the spatial statistics literature. In matching or weighting methods, geographic proximity has been incorporated to adjust for spatial confounding~\citep{keele2015enhancing,papadogeorgou2019adjusting,gao2022causal}. In these approaches, geographically proximate units are often preferentially matched under the assumption that geographic proximity captures shared unmeasured spatial factors~\citep{reich2021review}. A similar idea applies to network settings, where matching based on network-related covariates or summary measures of network structure has been proposed~\citep{zhang2023causal,awan2020almost}. However, close network proximity within matched sets may amplify autocorrelation of key variables, possibly worsening spurious associations due to network dependence. 
To our knowledge, there is a methodological gap in causal inference with observational network data where multiple variables are likely to exhibit network dependence.

\subsection{Notation and the null hypothesis of no causal effect}

Throughout this paper, we consider a network of $n$ units (nodes), with the adjacency matrix given by the $n\times n$ matrix $\bA =  [A_{ij}]$, where  $A_{ij} = 1$ indicates a tie (or edge) between units $i$ and $j$, and $A_{ij} = 0$ otherwise, for $i,j=1,2,\ldots, n$. Let $d_{ij}$ denote the geodesic distance (i.e., the length of the shortest path) between units $i$ and $j$. Throughout, we use ``geodesic distance" and ``network distance" interchangeably, although alternative distance measures on the network space may also be used to define $d_{ij}$. Let $Y_{i}$ denote an outcome variable, and let $Z_{i}$ denote a binary treatment assignment variable for unit $i$. 
Let $\bY = (Y_1,\cdots ,Y_n)$ and $\bZ = (Z_1,\cdots, Z_n)$ denote the corresponding collections of variables across the entire network. Let $\bX_i$ denote the $p\times 1$ covariate vector for unit $i$ and $\bX = (\bX_1,\cdots, \bX_n)^T$ denote the $n\times p$ covariate matrix.
The network is assumed to be fixed and given; thus the adjacency matrix is known. We say that a variable exhibits network dependence if its covariance structure depends on the network structure governed by the adjacency matrix. 

We introduce the potential outcomes framework to define the target null hypothesis and establish identification conditions~\citep{rubin1974estimating}. Let $\Omega = \{0,1\}^n$ denote the set of all possible treatment assignments for $\bZ$. 
For each unit $i$ and each treatment assignment vector $\bz \in \Omega$, let $Y_i(\bz)$ denote the potential outcome that would be observed for unit $i$ if the treatment assignment for all units were given by $\bz$. This representation allows for treatment interference~\citep{hudgens2008toward}.
We consider the following sharp null causal hypothesis of no causal effect:
\begin{eqnarray} \label{eq:null}
H_{0}: Y_i(\bz)= Y_i(\bz') \text{ for all } \bz, \bz' \in \Omega \text{ and }  i = 1,\cdots, n.
\end{eqnarray}
Rejecting $H_{0}$ may indicate the presence of a direct effect of a unit's treatment on its own outcome, an indirect effect on others' outcomes (treatment interference), or both. 

\subsection{Assumptions}

We first restrict the range of network dependence in treatment assignment using the network distance, $d_{ij}$. The condition below ultimately allows us to incorporate specific network proximity penalties into the matching procedures, as described in Section~\ref{sec:matching}. It also serves as a basis for establishing the validity of the proposed post-matching inference procedures introduced in Section~\ref{sec:inference}.

\begin{condition}[Local network dependence in treatment assignment]\label{assumption::local}
There exists $d_0\geq 1$ such that the treatment assignments of any two sets of units separated by more than $d_{0}$ are independent. Specifically, for any $I_1,I_2 \subseteq \{1,\ldots, n\}$ 
\[
\min_{i\in I_1,\,j\in I_2} d_{ij}>d_0
\quad\Longrightarrow\quad
\bZ_{I_1}\perp\bZ_{I_2}.
\]
\end{condition}
\noindent 
In particular, for a set $I\subseteq \{1,\cdots, n\}$, if  $d_{ij} >d_0$ for all distinct $i,j \in I$, repeated application of Condition~\ref{assumption::local} gives
$$\Pr(\bZ_I=\bz_I)=\prod_{i\in I}\Pr(Z_i=z_i).$$
Condition~\ref{assumption::local} motivates the construction of a set of treatment vectors whose corresponding units are pairwise separated by more than $d_{0}$, so that their assignments are mutually independent, thereby mitigating the impact of network dependence on treatment assignment. 
The local network dependence assumption is particularly natural in settings where interactions between units decay with distance. For example, consider the direct transmission process where units interact via network ties, and dependence propagates through these interactions~\citep{hays2010spatial,ogburn2018challenges,fowler2010cooperative,leung2022causal}. In such settings, the covariance in a given variable between two units becomes negligible (or zero) when their network distance exceeds a certain threshold, e.g., $d > d_{0} = 2$~\citep{dong2025disentangling}. Similar local dependence conditions can also be found in network and graphical-model literature~\citep{janisch2024berry, aronow2017estimating}.

For each unit $i$ and network distance $d$, define the $d$-neighborhood of unit $i$ as $\mathcal{N}_{d}(i)  = \{j \in \{1,\dots, n\}: d_{ij} \leq d\}$. The $d$-neighborhood of unit $i$ includes the unit itself, i.e., $i \in \mathcal{N}_d(i)$. Let $\bX_{\mathcal{N}_{d_0}(i)}$ denote the collection of covariate vectors for units in the $d_0$-neighborhood of unit $i$. Similarly, let $\bZ_{\mathcal{N}_{d_{0}}(i)}$ denote the corresponding vector of treatment assignments. 
In addition to consistency (i.e., $Y_{i} = Y_{i}(\bz)$ if $\bZ = \bz$, for any $i$), we impose the following causal identification assumptions for each unit $i$. 

\begin{assumption}[$d_{0}$-local interference]\label{assumption::interference}
$Y_{i}(\bz) = Y_{i}(\bz_{\mathcal{N}_{d_{0}}(i)})$ for any $\bz \in \Omega$.
\end{assumption}

\begin{assumption}[No unmeasured confounding]\label{assumption::ignorability}
$Y_{i}(\bz) \perp\!\!\!\perp \bZ_{\mathcal{N}_{d_{0}}(i)} \mid \bX_{\mathcal{N}_{d_0}(i)}$ for any 
$\bz \in \Omega$.
\end{assumption}
\begin{assumption}[Reducible individual treatment propensity]\label{assumption::inditreatment}
    $\Pr(Z_i = 1 | \bX_{\mathcal{N}_{d_0}(i)}) = \Pr(Z_i = 1|\bX_i)$
\end{assumption}
\noindent Assumption~\ref{assumption::interference} allows for interference while restricting its range to a finite ($d_{0}$) neighborhood. 
Assumption~\ref{assumption::ignorability} imposes a local ignorability condition, extending the standard no unmeasured confounding condition to settings with interference within the $d_0$-neighborhood. %
While complex covariance structures may exist within and across neighborhoods in $\bX$ and $\bY$ (in contrast to $\bZ$ under Condition~\ref{assumption::local}),  Assumption~\ref{assumption::ignorability} implies that, after conditioning on $\bX_{\mathcal{N}_{d_0}(i)}$, the residual variation in $\bZ_{\mathcal{N}_{d_{0}}(i)}$ is independent of the potential outcome $Y_{i}(\bz)$. Assumption~\ref{assumption::inditreatment} implies that, conditional on a unit's own covariates, $\bX_{\mathcal{N}_{d_0}(i)\setminus\{i\}}$ provides no additional information about that unit's treatment assignment. Assumptions concerning local interference, no unmeasured confounding, and reducible treatment propensity have been considered in the causal inference literature in network interference settings~\citep{forastiere2024causal,bhadra2025causal,forastiere2021identification,o2025local,lee2023estimating}.

\section{Dual-Penalty Matching Design}\label{sec:matching}

Our goal is to develop a matching method that achieves two objectives simultaneously: (i) reducing covariate dissimilarity between treated and control units to address confounding, and (ii) preventing network-proximate units from being matched together to mitigate spurious associations due to network dependence. 

\subsection{Distance measures with network proximity penalties}

We begin by defining distance measures to formalize the (dis)similarity between two units with respect to their network positions and covariates.
Network dissimilarity is defined based on the network distance $d_{ij}$, with smaller values indicating closer proximity within the network.  
As a covariate dissimilarity measure, without loss of generality, we introduce Mahalanobis distance, defined as $\delta_{ij} =  (\bX_i - \bX_j)^T \Sigma^{-1} (\bX_i - \bX_j)$, where $\Sigma$ is the pooled $p\times p$ variance-covariance matrix of the covariates across all units. Other distance metrics, such as propensity scores, may also be used to measure covariate dissimilarity.

Combining the two measures $d_{ij}$ and $\delta_{ij}$,  we propose a new measure  for matching: the Mahalanobis distance with a penalty on network proximity, defined as follows. 

\begin{equation} \label{eq:distance}
    \begin{split}
     D^{(\kappa)}_{ij} = \begin{cases}
   \delta_{ij}   &\text{  if  }  d_{ij} > \kappa\\
    \infty & \text{otherwise},
    \end{cases}
    \end{split}
\end{equation}
where $\kappa~(> 0)$ is a predefined threshold on network distance. 
This construction allows only pairs whose network distance exceeds $\kappa$ to be considered for matching; among such pairs, the distance measure depends solely on covariate dissimilarity.

\subsection{Dual-penalty matching design}\label{ssec:design}

We aim to construct a matching design $\mathcal{M}$ that matches all treated units to control units without discarding any observations. 
We formulate the matching problem as a mixed-integer program (MIP). MIP has been used to construct matched designs subject to study-specific balance and design
constraints~\citep{zubizarreta2012using,bennett2020building}.
Let $n_t$ and $n_c$ denote the total number of treated and control units, respectively, such that $n_t + n_c = n$.
Let $\mathcal{T} = \{t_1,t_2,\ldots,t_{n_t}\}$ and $\mathcal{C} = \{c_1,c_2,\ldots,c_{n_c}\}$ denote the sets of treated and control units, respectively, with $\mathcal{T} \cup \mathcal{C} = \{1,2,\ldots, n\}$.  For each pair $(t,c) \in \mathcal{T} \times \mathcal{C}$, define a binary decision variable $m_{tc}$, where $m_{tc} = 1$ if treated unit $t$ is matched to control unit $c$, and $m_{tc} = 0$ otherwise. 
The total network-proximity-penalized covariate dissimilarity across all matched treated--control pairs is given by
\begin{equation}
\label{eq:simpleobjective}
    \sum_{t\in \mathcal{T}}\sum_{c\in\mathcal{C}} D^{(\kappa)}_{tc}m_{tc}.
\end{equation}
This quantity serves as the objective function to be minimized and is similar to those used in standard optimal matching problems~\citep{gu1993comparison,stuart2008using}.
In our setting, a single penalty between treated-control pairs in $D^{(\kappa)}_{tc}$ is insufficient to separate all units in the network within a matched set. When a matched set consists of multiple treated or multiple control units, even if all between-treatment-arm pairs in the set satisfy $d_{tc}>\kappa$,
within-treatment-arm pairs (treated-treated/control-control pairs) in the same matched set may still be proximate in the network.

We impose additional constraints on within-treatment-arm pairs.
For each treated unit $t \in \mathcal{T}$, define the set of 
eligible controls as $\mathcal{C}_t = \{c\in \mathcal{C}: d_{tc}> \kappa\}$. Among those in $\mathcal{C}_t$, define the set of control pairs 
that cannot both be matched to the same treated unit $t$ as 
$\mathcal{P}_t  =  \{(c_1,c_2): c_1, c_2 \in \mathcal{C}_t,~d_{c_1c_2} \leq \kappa', c_1\neq c_2 \}$. 
Similarly, for each control unit $c \in \mathcal{C}$, define  
$\mathcal{T}_c = \{t\in \mathcal{T}: d_{tc} > \kappa\}$ and
$\mathcal{P}_c =  \{(t_1,t_2): t_1, t_2 \in \mathcal{T}_c,~d_{t_1 t_2 } \leq \kappa'', t_1\neq t_2 \}$. The thresholds $\kappa$, $\kappa'$, and $\kappa''$ need not be identical. For simplicity, we set $\kappa = \kappa'= \kappa''$ in the discussion below. 
Under Condition~\ref{assumption::local}, if $\kappa\geq d_0$ and every pair of units in a matched set has a network distance exceeding $\kappa$, treatment assignment within the matched set becomes independent. When $d_0$ is unknown, $\kappa$ should be interpreted as a design choice: larger values provide stronger network separation within matched sets but may reduce the number of feasible matches.

The proposed matching problem can be formulated as follows, with the objective of minimizing the total network-proximity-penalized covariate dissimilarity with additional constraints:
\begin{subequations}\label{eq:opt1}
\begin{align}
\min_{\mathcal{M}} \quad  &\sum_{t\in \mathcal{T}}\sum_{c\in \mathcal{C}} D^{(\kappa)}_{tc}\,m_{tc}  \label{eq:opt1a}\\
\text{subject to}\quad 
  & m_{tc_1} + m_{tc_2} \leq 1, 
  && \forall t\in \mathcal{T}, (c_1,c_2) \in \mathcal{P}_t \label{eq:opt1b}\\
  & m_{t_1c} + m_{t_2c} \leq 1, && \forall c\in \mathcal{C}, (t_1,t_2) \in \mathcal{P}_c  \label{eq:opt1c}\\
    & L_\mathcal{T} \leq \sum_{t\in \mathcal{T}} m_{tc} \leq U_\mathcal{T}, 
      && \forall c\in \mathcal{C}, \label{eq:opt1d}\\
  &L_\mathcal{C} \le \sum_{c\in \mathcal{C}} m_{tc} \leq U_\mathcal{C}, 
      && \forall t\in \mathcal{T}, \label{eq:opt1e}\\
      & m_{tc}\in\{0,1\}, 
      && \forall t \in \mathcal{T}, c\in \mathcal{C}, \label{eq:opt1f}
\end{align}
\end{subequations}
The objective~\eqref{eq:opt1a} minimizes the total network-proximity-penalized covariate dissimilarity across all between-treatment-arm pairs.
Constraints~\eqref{eq:opt1b}-\eqref{eq:opt1c} ensure that units from the same treatment arm cannot be placed in the same matched set if their pairwise distance is at most $\kappa$.
Specifically, even if two control units, $c_{1}$ and $c_{2}$, are each separated from a treated unit by a network distance greater than $\kappa$, they cannot both be included in the matched set containing $t$ if their network distance is at most $\kappa$. 
These constraints distinguish our matching method from conventional matching procedures, which account only for distances between treatment arms. We refer to our method as a \textit{dual-penalty} matching design because it penalizes network proximity both between and within treatment arms when forming matched sets. 
In contrast, we refer to a matching design that imposes a network penalty only between treatment arms as a \textit{single-penalty} design. 
Without constraints~\eqref{eq:opt1b}-\eqref{eq:opt1c}, the problem reduces to the conventional matching problem with the distance defined in~\eqref{eq:distance}~\citep{gu1993comparison,stuart2010matching}. 
Constraint~\eqref {eq:opt1f} is the standard integrality condition that prevents fractional assignments. 

Constraints~\eqref{eq:opt1d}-\eqref{eq:opt1e} regulate matched-set sizes, where $L_\mathcal{C}$ and $U_\mathcal{C}$ denote the lower and upper bounds on the number of controls matched to each treated unit, and $L_\mathcal{T}$ and $U_\mathcal{T}$ denote the corresponding bounds on the number of treated units matched to each control. For example, setting $(L_\mathcal{C},U_\mathcal{C}) = (m,m)$ and $(L_\mathcal{T}, U_\mathcal{T}) = (0,1)$ yields an $m$:1 matching design. 
In what follows, we set $(L_\mathcal{C},U_\mathcal{C}) = (1,\infty)$ and $ (L_\mathcal{T},U_\mathcal{T}) = (1,\infty)$, which yields a full matching design~\citep{hansen2004full}. Under this design, each matched set contains either one treated unit and one or more control units, or one control unit and one or more treated units.  
We exclude the many-to-many design because it can always be partitioned into a full matching design with a smaller total covariate dissimilarity (see Section 5.2 in~\cite{zubizarreta2023handbook}).

\subsection{Computational complexity and implementation}
\label{ssec:computation}
The formulation in~\eqref{eq:opt1} involves $n_tn_c$ decision variables and a large number of within-treatment-arm constraints. In the worst case, \eqref{eq:opt1b} and \eqref{eq:opt1c} impose $n_t\binom{n_c}{2}$ and $n_c\binom{n_t}{2}$ constraints, respectively, yielding a total of $O\{n_tn_c(n_t+n_c)\}$ constraints.
Although solving such optimization problems can be computationally challenging~\citep{oncan2013minimum,agrawal2020parameterized}, MIP-based matching formulations are often computationally tractable in practice~\citep{bennett2020building}.
In the real data application presented in Section~\ref{sec:application}, the MIP problem of the dual-penalty matching on a network of size $n=2,347$ and $\kappa = 2$ 
is solved in 128.09 seconds. 
In the \texttt{R} package \texttt{netmatchRI}, we solve the MIP problem using the open-source HiGHS~\citep{huangfu2018parallelizing}.  We also provide alternative implementations using 
the open-source GNU Linear Programming Kit (GLPK)~\citep{makhorin2008glpk} and the commercial Gurobi Optimizer~\citep{gurobi} (license required).

\subsection{Illustrative toy example}\label{ssec::illustDesign}

\begin{figure}[!ht]
\centering

\tikzset{
 treated/.style={
   circle,
   draw=black,
   minimum size=14pt,
   inner sep=0pt,
   font=\fontsize{6.8}{6.8}\selectfont\bfseries,
   preaction={fill=#1},
   pattern=north east lines,
   pattern color=black!70
 },
 control/.style={
   circle,
   draw=black,
   minimum size=14pt,
   inner sep=0pt,
   font=\fontsize{6.8}{6.8}\selectfont\bfseries,
   fill=#1
 },
 network label/.style={font=\bfseries, text depth=0pt, text height=1em},
 net edge/.style={draw=black!55, line width=0.42pt}
}

\def\cOne{green!30}
\def\cTwo{blue!28}
\def\cThree{yellow!28}
\def\cFour{orange!35}
\def\cFive{red!32}
\def\cSix{violet!22}

\newcommand{\covnetcoords}{
  \coordinate (p1)  at (0.0,4.8);
  \coordinate (p2)  at (0.7,3.45);
  \coordinate (p3)  at (0.15,2.10);
  \coordinate (p4)  at (-1.10,1.55);
  \coordinate (p5)  at (1.30,1.15);
  \coordinate (p6)  at (1.85,2.70);
  \coordinate (p7)  at (2.35,1.75);
  \coordinate (p8)  at (2.00,0.35);
  \coordinate (p9)  at (3.35,0.65);
  \coordinate (p10) at (3.15,2.45);
  \coordinate (p11) at (4.45,3.15);
  \coordinate (p12) at (5.20,2.15);
  \coordinate (p13) at (5.60,3.75);
  \coordinate (p14) at (6.45,1.95);
  \coordinate (p15) at (5.25,0.80);
  \coordinate (p16) at (4.10,0.05);
}

\newcommand{\covnetedges}{
  \draw[net edge] (n1)  -- (n2);
  \draw[net edge] (n2)  -- (n3);
  \draw[net edge] (n3)  -- (n4);
  \draw[net edge] (n3)  -- (n5);
  \draw[net edge] (n2)  -- (n6);
  \draw[net edge] (n5)  -- (n6);
  \draw[net edge] (n5)  -- (n8);
  \draw[net edge] (n6)  -- (n7);
  \draw[net edge] (n6)  -- (n10);
  \draw[net edge] (n7)  -- (n8);
  \draw[net edge] (n7)  -- (n10);
  \draw[net edge] (n8)  -- (n9);
  \draw[net edge] (n9)  -- (n16);
  \draw[net edge] (n10) -- (n11);
  \draw[net edge] (n10) -- (n12);
  \draw[net edge] (n10) -- (n16);
  \draw[net edge] (n11) -- (n12);
  \draw[net edge] (n12) -- (n13);
  \draw[net edge] (n12) -- (n14);
  \draw[net edge] (n12) -- (n15);
  \draw[net edge] (n15) -- (n16);
}

\begin{subfigure}[t]{0.32\textwidth}
\centering
\resizebox{0.95\linewidth}{!}{%
\begin{tikzpicture}[x=0.92cm,y=0.92cm,-]
\covnetcoords
\node[network label, anchor=base] at (2.70,5.25) {Covariates only};

\node[control=\cOne]   (n1)  at (p1)  {0.1};
\node[control=\cTwo]   (n2)  at (p2)  {1.4};
\node[control=\cTwo]   (n3)  at (p3)  {0.2};
\node[control=\cTwo]   (n4)  at (p4)  {0.2};
\node[control=\cTwo]   (n5)  at (p5)  {1.9};
\node[control=\cTwo]   (n6)  at (p6)  {2.0};
\node[control=\cFive]  (n7)  at (p7)  {3.3};
\node[treated=\cTwo]   (n8)  at (p8)  {2.0};
\node[control=\cThree] (n9)  at (p9)  {2.1};
\node[treated=\cFour]  (n10) at (p10) {3.2};
\node[control=\cFour]  (n11) at (p11) {3.3};
\node[treated=\cFive]  (n12) at (p12) {3.4};
\node[control=\cSix]   (n13) at (p13) {2.9};
\node[treated=\cSix]   (n14) at (p14) {2.7};
\node[treated=\cOne]   (n15) at (p15) {2.4};
\node[treated=\cThree] (n16) at (p16) {2.4};

\covnetedges
\end{tikzpicture}%
}
\caption{}
\label{fig:penal:only}
\end{subfigure}
\hfill
\begin{subfigure}[t]{0.32\textwidth}
\centering
\resizebox{0.95\linewidth}{!}{%
\begin{tikzpicture}[x=0.92cm,y=0.92cm,-]
\covnetcoords
\node[network label, anchor=base] at (2.70,5.25) {Single penalty};

\node[control=\cTwo]   (n1)  at (p1)  {0.1};
\node[control=\cTwo]   (n2)  at (p2)  {1.4};
\node[control=\cFour]   (n3)  at (p3)  {0.2};
\node[control=\cTwo]   (n4)  at (p4)  {0.2};
\node[control=\cOne]   (n5)  at (p5)  {1.9};
\node[control=\cOne]   (n6)  at (p6)  {2.0};
\node[control=\cSix]  (n7)  at (p7)  {3.3};
\node[treated=\cTwo]   (n8)  at (p8)  {2.0};
\node[control=\cFive]   (n9)  at (p9)  {2.1};
\node[treated=\cFour]  (n10) at (p10) {3.2};
\node[control=\cTwo] (n11) at (p11) {3.3};
\node[treated=\cFive]  (n12) at (p12) {3.4};
\node[control=\cThree]  (n13) at (p13) {2.9};
\node[treated=\cSix]   (n14) at (p14) {2.7};
\node[treated=\cOne]   (n15) at (p15) {2.4};
\node[treated=\cThree] (n16) at (p16) {2.4};

\covnetedges
\end{tikzpicture}%
}
\caption{}
\label{fig:penal:single}
\end{subfigure}
\hfill
\begin{subfigure}[t]{0.32\textwidth}
\centering
\resizebox{0.95\linewidth}{!}{%
\begin{tikzpicture}[x=0.92cm,y=0.92cm,-]
\covnetcoords
\node[network label, anchor=base] at (2.70,5.25) {Dual penalty};

\node[control=\cThree] (n1)  at (p1)  {0.1};
\node[control=\cTwo]   (n2)  at (p2)  {1.4};
\node[control=\cFour]   (n3)  at (p3)  {0.2};
\node[control=\cOne]   (n4)  at (p4)  {0.2};
\node[control=\cThree] (n5)  at (p5)  {1.9};
\node[control=\cOne]   (n6)  at (p6)  {2.0};
\node[control=\cSix]  (n7)  at (p7)  {3.3};
\node[treated=\cTwo]   (n8)  at (p8)  {2.0};
\node[control=\cFive]   (n9)  at (p9)  {2.1};
\node[treated=\cFour]  (n10) at (p10) {3.2};
\node[control=\cTwo] (n11) at (p11) {3.3};
\node[treated=\cFive]  (n12) at (p12) {3.4};
\node[control=\cThree]  (n13) at (p13) {2.9};
\node[treated=\cSix]   (n14) at (p14) {2.7};
\node[treated=\cOne]   (n15) at (p15) {2.4};
\node[treated=\cThree] (n16) at (p16) {2.4};

\covnetedges
\end{tikzpicture}%
}
\caption{}
\label{fig:penal:double}
\end{subfigure}

\vspace{0.45em}
\resizebox{0.9\linewidth}{!}{%
\begin{tikzpicture}[x=2cm,y=2cm]
\covnetcoords
\node[treated=white] at (0,0) {};   \node[anchor=west] at (0.18,0) {\small Treated};
\node[control=white] at (1.6,0) {}; \node[anchor=west] at (1.78,0) {\small Control};
\node[anchor=west]   at (3.3,0) {\small Subclass:};
\node[control=\cOne]   at (4.4,0) {}; \node[anchor=west] at (4.55,0) {\small 1};
\node[control=\cTwo]   at (5.2,0) {}; \node[anchor=west] at (5.35,0) {\small 2};
\node[control=\cThree] at (6.0,0) {}; \node[anchor=west] at (6.15,0) {\small 3};
\node[control=\cFour]  at (6.8,0) {}; \node[anchor=west] at (6.95,0) {\small 4};
\node[control=\cFive]  at (7.6,0) {}; \node[anchor=west] at (7.75,0) {\small 5};
\node[control=\cSix]   at (8.4,0) {}; \node[anchor=west] at (8.55,0) {\small 6};
\end{tikzpicture}}

\caption{Comparison of three matching designs with covariate values shown inside the nodes. Treated units are marked with diagonal shading, and control units are unshaded. Colors indicate the matched subclasses defined under each design. Figure~\ref{Figure::toyour}\subref{fig:penal:only} shows the covariates-only matching, Figure~\ref{Figure::toyour}\subref{fig:penal:single} shows the single-penalty matching, and Figure~\ref{Figure::toyour}\subref{fig:penal:double} shows the dual-penalty matching.}
\label{Figure::toyour}
\end{figure}
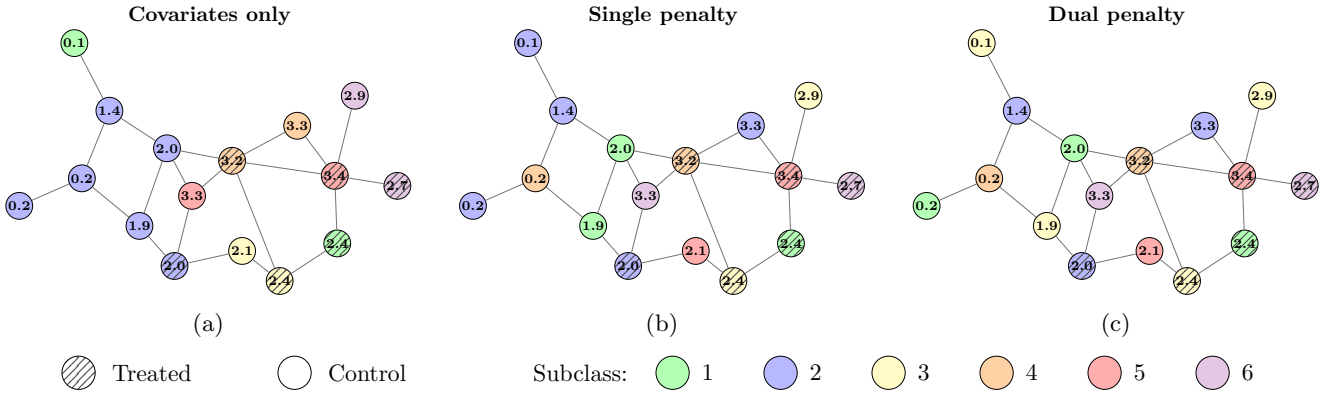

To demonstrate the role of constraints in~\eqref{eq:distance},~\eqref{eq:opt1b} and~\eqref{eq:opt1c} in the matching formulation, we present a toy example of $n=16$ units in Figure~\ref{Figure::toyour}. Baseline covariates and treatment assignments are arbitrarily assigned and exhibit network dependence (e.g., nearby units tend to have similar covariate values and treatment assignments). In each panel of Figure~\ref{Figure::toyour}, covariate values are shown within each node; treated units are indicated by diagonal shading, and control units are unshaded. 
We consider three matching designs: (1) covariate-only matching, which constructs matched sets by minimizing total $\delta_{tc}$ for treated--control pairs without any penalties on network proximity, (2) single-penalty matching, which minimizes total $D^{(\kappa)}_{tc}$ but omits constraints~\eqref{eq:opt1b}-\eqref{eq:opt1c}
and (3) dual-penalty matching, which minimizes total $D^{(\kappa)}_{tc}$ with constraints~\eqref{eq:opt1b}-\eqref{eq:opt1c}.
We set $\kappa =2$, corresponding to the range of local dependence in the toy example under Condition~\ref{assumption::local}, where $d_0 =2$.

Figure~\ref{Figure::toyour} presents the matching result under each design, with different colors indicating different matched sets (subclasses). 
Covariate-only matching (Figure~\ref{Figure::toyour}\subref{fig:penal:only}) produces several matched sets containing units within a network distance of two. Although this approach minimizes covariate dissimilarity, it may result in matched sets that are easily susceptible to spurious associations in subsequent analyses due to network dependence. 
In contrast, single-penalty matching (Figure~\ref{Figure::toyour}\subref{fig:penal:single}) separates some pairs within a network distance of two or less, but only across different treatment arms. 
Under the proposed dual-penalty matching (Figure~\ref{Figure::toyour}\subref{fig:penal:double}), the pairwise network distance between \textit{any} pair of units within a matched set exceeds two, rendering their treatment assignments mutually independent when $\kappa$ is chosen with $\kappa \geq d_{0}$.

\section{Randomization-based inference with network-dependent matched sets}\label{sec:inference}

In this section, conditional on the matched design $\mathcal{M}$, we develop valid randomization-based inference procedures, establish conditions under which the test statistics are asymptotically normal, and construct bounds on the covariances of the test statistics across matched sets.

A matching design $\mathcal{M}$ partitions $n$ units into $S$ matched sets, where each matched set $s \in \{1,\ldots,S\}$ contains $n_{st}$ treated and $n_{sc}$ control units ($n_s = n_{st}+n_{sc}$).  Let $Z_{si}$, $Y_{si}$  and $\bX_{si}$ denote the treatment assignment, outcome, and $p\times 1$ vector of covariates  for unit $i$ in matched set $s$, respectively ($s=1,\ldots, S,~i=1,\ldots, n_{s}$).
Let $\bZ_s = (Z_{s1},\ldots,Z_{sn_s})$ denote the treatment assignment vector for matched set $s$. 
Let $\bZ^\mathcal{M} = (\bZ_1,\cdots, \bZ_S)$ denote the full treatment assignment vector after matching. Since $\bZ^\mathcal{M}$ is a relabeling of the original treatment assignment $\bZ$, we suppress the superscript $\mathcal{M}$ and write $\bZ$ hereafter.
Let $\Omega_\mathcal{M}= \prod_{s=1}^{S}\Omega_s$ denote the set of possible treatment assignments for $\bZ$ conditional on  $\mathcal{M}$,  where 
$\Omega_s = \{\bz_s \in \{0,1\}^{n_s}: \textstyle\sum^{n_s}_{i=1} z_{si} = n_{st}\}$ is the set of all possible treatment assignments within matched set $s$.
For a treatment assignment vector $\bz \in \Omega_{\mathcal{M}}$, let 
$Y_{si}(\bz)$ be the potential outcome of unit $i$ in matched set $s$  
under $\bz$. Define  $\bY_s(\bz) = (Y_{s1}(\bz),\ldots,Y_{sn_s}(\bz))$ as the vector of potential outcomes in matched set $s$, and $\bY(\bz) = (\bY_1(\bz),\cdots, \bY_S(\bz))$ as the vector of potential outcomes for all matched sets.

Conditional on  $\mathcal{M}$, we now test the null hypothesis $H_{0}$ in~\eqref{eq:null} through the following matching-specific null hypothesis $H_{0, \mathcal{M}}$:
\begin{eqnarray} \label{eq:nullM}
H_{0,\mathcal{M}}: Y_{si}(\bz) = Y_{si}(\bz') \quad 
\text{for all } s,\, i,\, \bz,\bz' \in \Omega_{\mathcal{M}}.
\end{eqnarray}
\noindent We use the randomization inference framework~\citep{rosenbaum2002observational}, in which the potential outcomes are treated as fixed and the treatment assignment mechanism is the source of randomness in the test statistics. In our setting, without guaranteeing independence within and across $\bZ_s$'s, we develop an asymptotic randomization-based test that uses a normal approximation to the randomization distribution of the test statistic.   

\subsection{Nonparametric test statistic and its asymptotic normality}\label{sec::testandRI}

Let $T(\bZ,\bY)$ denote a test statistic of interest for testing~\eqref{eq:nullM}. Under  $H_{0,\mathcal{M}}$, the potential outcomes are invariant to the treatment assignment for all $\bz \in \Omega_{\mathcal{M}}$.
We consider a weighted sum of Mann-Whitney U statistics~\citep{lehmann2006nonparametrics}:
\begin{align*} \label{eq:test}
    T(\bZ, \bY) &= \sum^S_{s=1}w_s U_{s}(\bZ), \quad w_s =\dfrac{1}{n_s + 1},
\end{align*}
where $U_s(\bZ) =\sum^{n_s}_{i=1}\sum^{n_s}_{j=1} Z_{si}(1-Z_{sj})f_{sij\bZ}$, 
$f_{sij\bZ} =\mathbf{1}\{Y_{si} (\bZ)> Y_{sj}(\bZ)\}$, with $\mathbf{1}\{\cdot\}$ denoting an indicator function, and $w_s$ denoting the weight assigned to $U_s(\bZ)$ for matched set $s$.  
This nonparametric statistic has been used in randomization inference under interference~\citep{rosenbaum2007interference} and in group-randomized trials~\citep{zhang2012powerful,small2008randomization,kawaguchi2011stratified}. 
For notational simplicity, we write $U_s$ for $U_s(\bZ)$ hereafter because $U_{s}(\bZ)$ depends on $\bZ$ only through $\bZ_{s}$. Let $\mu_s$ and $\sigma_{ss}$ be the mean and variance of $U_s$, respectively, and let $\sigma_{sl} = \text{Cov}(U_s, U_l)$ denote the covariance between the test statistics from two matched sets $s$ and $l$. The mean and variance of $T(\bZ,\bY)$ can be written as:
\begin{equation}\label{eq:EVofT}
    \mu_{T}:=\E[T(\bZ, \bY)] =\sum^S_{s=1} w_s\mu_s, \quad  \sigma^2_T := \V[T(\bZ, \bY)] = \sum^S_{s=1}w^2_s\sigma_{ss} + \sum^{S}_{s< l}2w_sw_l \sigma_{sl},
\end{equation}

First, to establish the distribution of each $U_s$, we impose the following condition, which ensures that every pair of units within a matched set $s$ is separated by a network distance greater than $\kappa$.
\begin{condition}[Network-distance  separation within matched sets]\label{condition::separation}
The matched design $\mathcal{M}$ is constructed such that every pair of units within each matched set has a  network distance $d_{ij} > d_{0}$.
\end{condition}
\noindent Condition~\ref{condition::separation} is satisfied when the proposed dual-penalty matching method is used with $\kappa \geq d_{0}$. 
Then combined with Condition~\ref{assumption::local}, the components of $\bZ_{s}$ behave as if they were randomly assigned. 
\begin{corollary} \label{cor::ignorability}
Under Conditions~\ref{assumption::local}-\ref{condition::separation}, and Assumptions~\ref{assumption::interference}-\ref{assumption::inditreatment}, for each unit $i$ in matched set $s$,
$$Y_{si}(\bz') \perp\!\!\!\perp Z_{si} \,\big|\, \bX_{si}\quad \text{for any } \bz' \in \Omega_\mathcal{M}.$$
\end{corollary}
\noindent 
Corollary~\ref{cor::ignorability} implies that Assumption~\ref{assumption::ignorability} reduces to individual treatment ignorability, allowing us to use unit-level covariates in matching to address observed confounding.

\begin{condition}\label{assumption::exact}
For each matched set $s$, there exists $p_s\in(0,1)$, such that all units in the matched set have a common individual treatment propensity, $p_{s}$. 
\end{condition}
\noindent This common treatment probability condition holds under exact matching on covariates~\citep{imai2008variance}.  Corollary~\ref{cor::ignorability} and Condition~\ref{assumption::exact} together enable us to treat each matched set as a randomized experiment with fixed numbers of treated and control units. The following lemma characterizes the distribution of the matched-set-specific test statistic $U_{s}$ and establishes that its moments are fixed. 
\begin{lemma}\label{lemma::moments}
Suppose that Conditions~\ref{assumption::local}-\ref{assumption::exact} and Assumptions~\ref{assumption::interference}-\ref{assumption::inditreatment} hold. If there are no tied outcomes within matched sets, 
\begin{equation*}\label{eq:uniformU}
    \Pr(U_s = k) = \frac{c(n_{st}, n_{sc}, k)}{\binom{n_{st}+n_{sc}}{n_{st}}}, \quad k = 0, 1, \ldots, n_{st}n_{sc},
\end{equation*}
where 
$c(n_{st}, n_{sc}, k) = \left|\left\{\bz_s\in\Omega_s:U_s= k\right\}\right|$ counts the assignments in $\Omega_s$ that yield $U_s=k$.
Moreover,
\begin{eqnarray}\label{eq:mean}
\mu_s = \frac{n_{st}n_{sc}}{2}, \quad \sigma_{ss} =\frac{n_{st}n_{sc}(n_s+1)}{12}.
\end{eqnarray}
\end{lemma}
\noindent 
The proofs of Corollary~\ref{cor::ignorability} and Lemma~\ref{lemma::moments} are provided in Supplementary Material~\ref{apdx:proof}.
We next show that, under certain regularity conditions, $T(\bZ,\bY)$ is asymptotically normal with mean $\mu_T$ and variance $\sigma^2_T$ as in~\eqref{eq:EVofT}, without requiring explicit specification of the covariances $\sigma_{sl}$.

\begin{theorem}[Asymptotic normality of the test statistic]\label{theorem::clt}
Suppose the conditions of Lemma~\ref{lemma::moments} and Conditions~\ref{condition::boundedmatchedsets}--\ref{condition::nondevaraince} in Supplementary Material~\ref{apdx:cltproof} hold.
 Then, as $S\to\infty$,
\[
\frac{T(\bZ,\bY)-\mu_T}{\sigma_T}\xrightarrow{d}\mathcal{N}(0,1).
\]
\end{theorem}
\noindent Theorem~\ref{theorem::clt} justifies that the randomization distribution of $T(\bZ,\bY)$ can be approximated by a normal distribution with mean $\mu_T$ and variance $\sigma^2_{T}$ when the number of matched sets is large. The proof of Theorem~\ref{theorem::clt} is provided in Supplementary Material~\ref{apdx:cltproof}, where we impose regularity conditions on network structures and variables; similar conditions can be found in other relevant literature~\citep{chandrasekhar2023general, aronow2017estimating, gao2025causal}. 

Supplementary Material~\ref{apdx:permutation} presents an alternative inference approach for settings in which the treatment assignment mechanism $\Pr(\bZ)$ can be specified using the known dependence structure of $\bZ$. In this case, the randomization distribution of $T(\bZ,\bY)$ can be approximated using Monte Carlo draws of $\bZ$ from $\Pr(\bZ)$. However, misspecification of the treatment assignment mechanism may invalidate the resulting inference. In contrast, inference based on the normal approximation may rely on bounds for the covariances $\sigma_{sl}$'s, rather than requiring a fully specified $\Pr(\bZ)$, as detailed in the next subsection.

\subsection{Bounds on cross-matched-set covariances}\label{sec:designbound}

We aim to specify the maximum covariances $\sigma_{sl}$'s using the matched-set-specific distribution implied by the matched design. Under certain conditions, the proposed dual-penalty matching method not only enables valid causal comparisons within matched sets but also induces the fixed marginal distributions of the set-specific test statistics established in Lemma~\ref{lemma::moments}. These distributions can then be used to construct bounds on the covariances across matched sets. 

\begin{theorem}[Design-based upper bound on cross-matched-set covariance]\label{theorem::bound} Suppose the conditions of Lemma~\ref{lemma::moments} hold. For distinct matched sets $s$ and $l$, let $K_s=n_{st}n_{sc}$ and $K_l=n_{lt}n_{lc}$, and let $F_s$ and $F_l$ denote the marginal cumulative distribution functions (CDFs) of $U_s$ and $U_l$, respectively. Define the Fr\'echet--Hoeffding upper bound $F^M_{s,l}(a,b)=\min\{F_s(a),F_l(b)\}$, with $F_s(-1)=F_l(-1)=0$, and let $p^M_{ab}=F^M_{s,l}(a,b)-F^M_{s,l}(a-1,b)-F^M_{s,l}(a,b-1)+F^M_{s,l}(a-1,b-1)$. Then \begin{equation}\label{eq:designboundM}\sigma_{sl}=\operatorname{Cov}(U_s,U_l)\leq M_{sl}:=\sum_{a=0}^{K_s}\sum_{b=0}^{K_l}ab\,p^M_{ab}-\mu_s\mu_l.\end{equation} The bound is sharp and is attained by the joint distribution with CDF $F^M_{s,l}$. \end{theorem}

\noindent The bound $M_{sl}$ is design-based because it depends only on the fixed marginal distributions of the set-specific test statistics as a result of matched design.
Replacing $\sigma_{sl}$'s in $\V[T(\bZ,\bY)]$ with $M_{sl}$'s yields a design-based upper bound on $\sigma^2_T$,
\begin{equation}\label{eq:designbound}\sigma^2_T\le\mathbb{V}_M[T(\bZ,\bY)]:=\sum_{s=1}^S w_s^2\sigma_{ss}+\sum_{\substack{s<l\\ l\in\mathcal{I}_{\kappa}(s)}} 2 w_s w_l\,M_{sl},
\end{equation} 
where $\mathcal{I}_{\kappa}(s) = \{l: \min\{d_{ij}, i \in s, j\in l\} \le \kappa\}$ denotes the collection of matched sets whose treatment assignment may depend on those in matched set $s$. The proof of Theorem~\ref{theorem::bound} is provided in Supplementary Material~\ref{ssec:proof2}.
\noindent Because $\sigma_T^2\le\mathbb{V}_M[T(\bZ,\bY)]$ in~\eqref{eq:designbound}, using $\mathbb{V}_M[T(\bZ,\bY)]$ rather than $\sigma_T^2$ to standardize the test statistic yields a test that asymptotically controls the Type I error rate as the number of matched sets increases. However, inference based on the design-based upper bound may be asymptotically conservative. 

By imposing structural assumptions on network dependence, we may tighten $M_{sl}$ and thereby obtain smaller covariance bounds. In many network settings, dependence between two units is often expected to weaken as network distance between them increases~\citep{fowler2010cooperative,milgram1967small, leung2022causal}. We incorporate this observation into the covariance bound. Define $h_{s,l} := \min \{d_{ij},~i \in s,~j \in l\}$ as the minimum network distance between any unit in matched set $s$ and any unit in matched set $l$. 
Then for two matched sets $s$ and $l$ with $l \in \mathcal{I}_{\kappa}(s)$, define
\begin{equation}\label{eq:modelboundM}
M^*_{sl}(\eta,\rho)=\eta\,\rho^{\,h_{s,l}-1}\,M_{sl},
\end{equation}
where $M_{sl}$ is the design-based bound in~\eqref{eq:designboundM}, $\eta\in[0,1]$ controls the overall magnitude of the covariance bound relative to the design-based bound, and $\rho\in[0,1]$ determines how rapidly this bound decays as the minimum network distance between the two sets increases. In particular, setting $\eta = 0$ eliminates all cross-matched-set covariance contributions from the variance bound. 
If $\eta = 1$, $M^*_{sl}(\eta,\rho)$ equals $M_{sl}$ when $h_{s,l} = 1$. Smaller values of $\rho$ imply that the covariance bound decays more rapidly with the minimum network distance between matched sets, whereas $\rho = 1$ corresponds to no decay.
When $\eta = \rho = 1$, $M^*_{sl}(\eta,\rho)$ reduces to $M_{sl}$ for all $h_{s,l}$. 
Replacing the cross-matched-set covariances in~\eqref{eq:designbound} with the bounds in~\eqref{eq:modelboundM}
yields
\begin{equation}\label{eq:modelbound}
    \V_{\text{M}^*(\eta,\rho)}[T(\bZ,\bY)]:=\sum_{s=1}^S w_s^2\sigma_{ss}+\sum_{\substack{s<l\\ l\in\mathcal{I}_{\kappa}(s)}} 2 w_s w_l\,M^*_{sl}(\eta,\rho).
\end{equation}
\noindent When the true covariance $\sigma_{sl}$ does not exceed $M^*_{sl}(\eta,\rho)$ for every $s$ and every $l\in\mathcal{I}_{\kappa}(s)$, $\V_{\text{M}^*(\eta,\rho)}[T(\bZ,\bY)]$ in \eqref{eq:modelbound} provides a valid upper bound on $\sigma^2_T$.  Therefore, the resulting test asymptotically controls the Type I error as the number of matched sets increases. Without knowledge of the true value of $(\eta, \rho)$, some specifications may underestimate the true covariances and thereby lead to invalid inference. 
Supplementary Material~\ref{ssec:sensitivity} explains how the parameter pair $(\eta, \rho)$ can be interpreted as sensitivity parameters governing the covariance bounds. 

\section{Simulation studies} \label{sec:simulation}

Through simulation studies, we aim to evaluate the effectiveness of the proposed dual-penalty matching method and randomization-based inference procedures in addressing spurious associations due to network dependence.

\subsection{Simulation settings} \label{ssec:simsetting}

We generate a fixed undirected network of $n=300$ units from a stochastic block model~\citep{karrer2011stochastic}, with four equal-size blocks. The within- and between-block tie probabilities are $p_{\text{in}}=0.19$ and $p_{\text{out}}=0.003$, respectively. The resulting network has an average degree of $14.9$ and an average network distance of $2.9$. Let $\bA$ denote the adjacency matrix and $\bW$ its row-standardized version.
We generate the outcome $\bY$, covariates $\bX$, and treatment assignments $\bZ$ using the common $n\times n$ variance-covariance matrix $\bV = (\alpha_{1}\bW + \alpha_{2}\bI_{n})(\alpha_{1}\bW + \alpha_{2}\bI_n)^T$, where $\bI_{n}$ is the $n \times n$ identity matrix, $\alpha_1$ controls the strength of network-induced dependence through $\bW$, and $\alpha_2$ captures network-independent, node-specific variation.
This specification induces dependence among units separated by a network distance at most two ($d_0 = 2$) and is motivated by covariance structures arising from direct transmission processes (e.g.,~\cite{aral2017exercise,banerjee2013diffusion}).
We consider three dependence levels: \emph{independent} ($\alpha_1=0,\alpha_2=1$), \emph{moderate} ($\alpha_1=0.7,\alpha_2=0.3$), and \emph{strong} ($\alpha_1=0.9,\alpha_2=0.1$). We set the causal effect of $\bZ$ on $\bY$ to $\beta_z = 0$ under the null and vary $\beta_z$ in the power analysis. 
Further details are provided in Supplementary Material~\ref{apdx::sim:setting}.

\subsection{Performance of matching methods} \label{sec:sim_results1}

\begin{figure}[!ht]
\centering
   \begin{subfigure}[t]{0.42\textwidth}
   \centering\includegraphics[width=\linewidth]{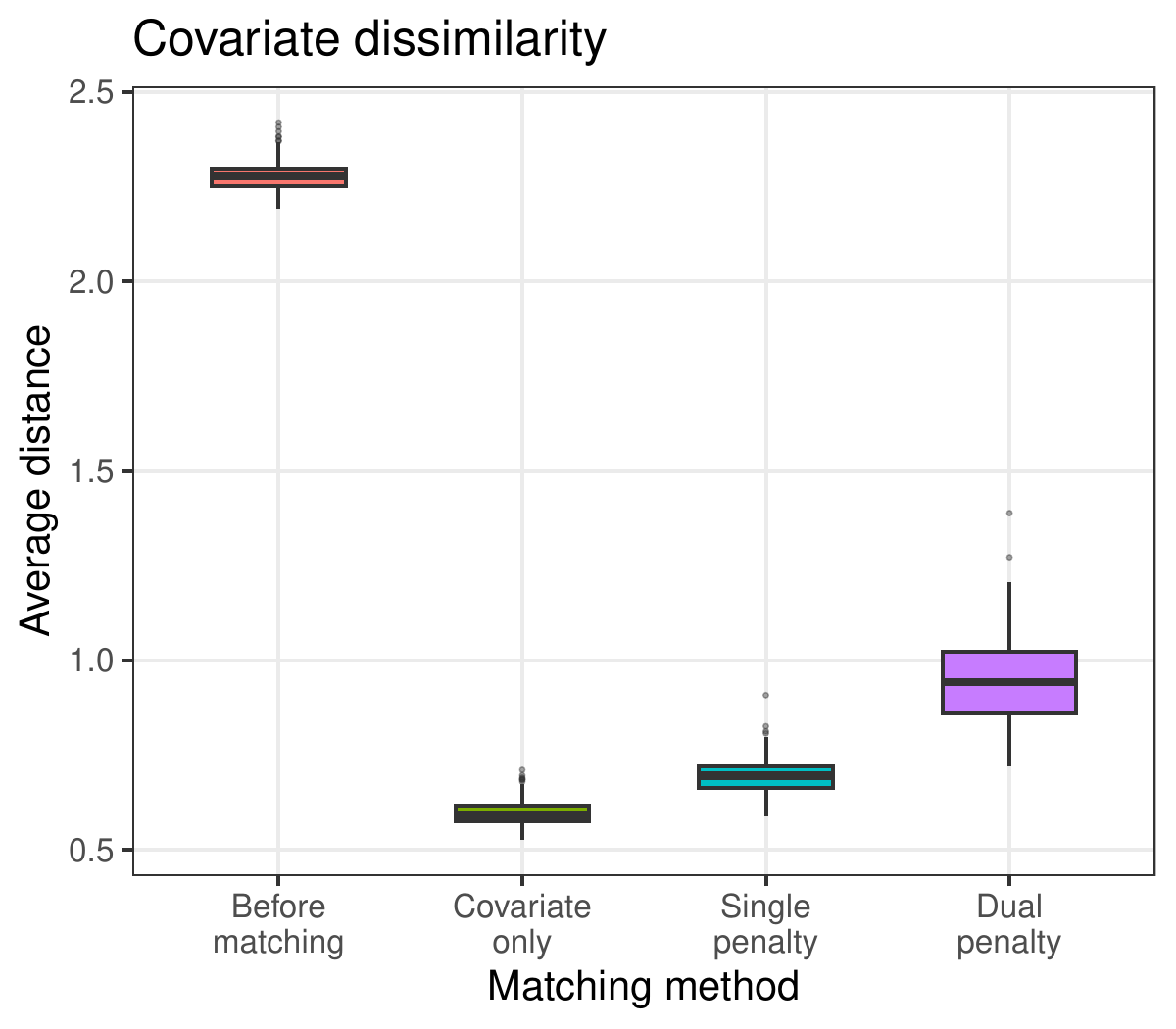}
    \caption{Average covariate dissimilarity}
     \label{fig:cov}
   \end{subfigure}
   \begin{subfigure}[t]{0.42\textwidth}
    \centering
    \includegraphics[width=\linewidth]{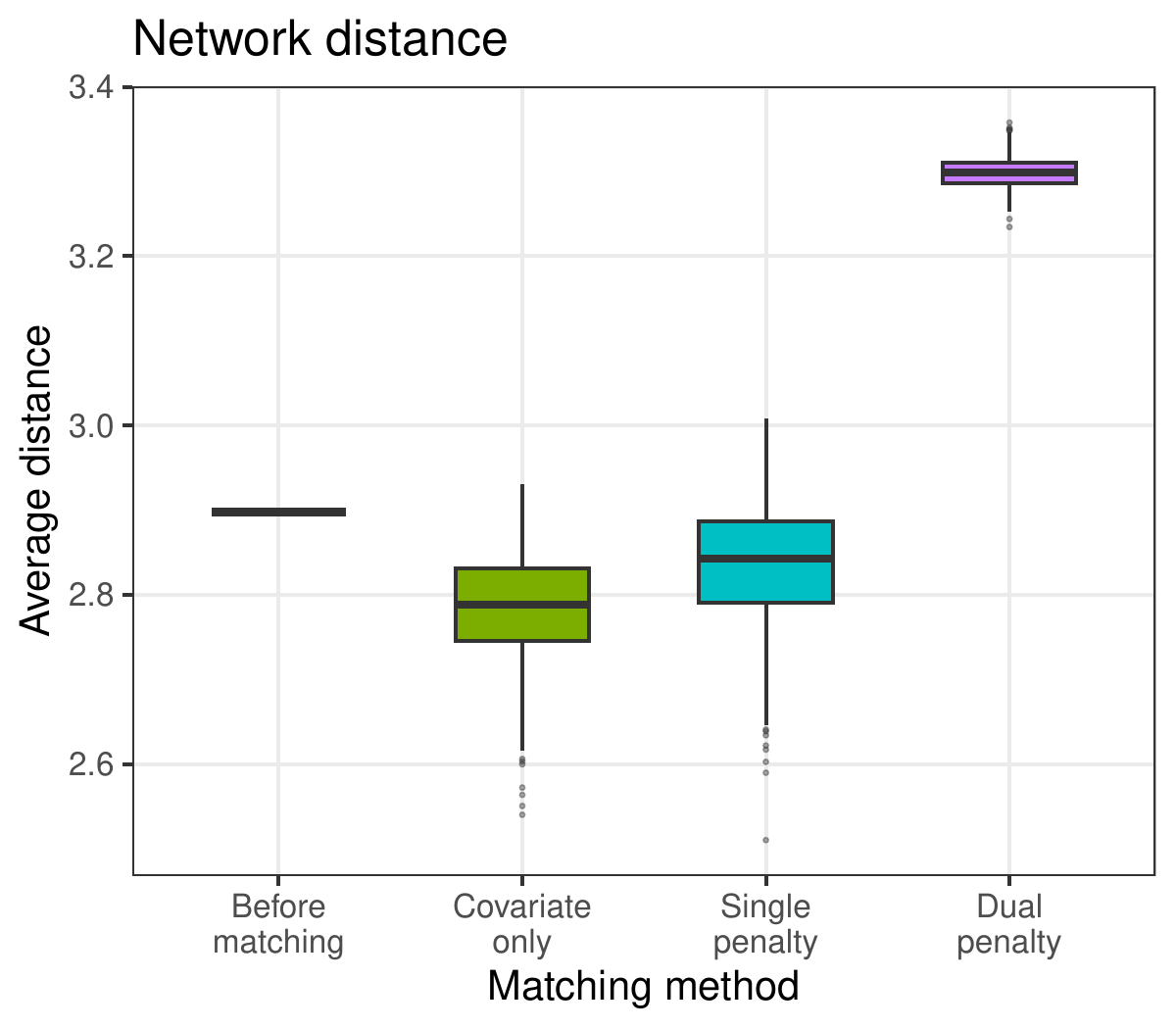}
    \caption{Average network distance}
    \label{fig:net-avg}
  \end{subfigure}
  \caption{Matching diagnostics under strong network dependence and null ($\beta_z = 0$), over 500 replications:  (a) distributions of the average covariate dissimilarity, measured by $\delta_{ij}$, before and after matching; and (b) distributions of the average within-set network distance, $d_{ij}$, before and after matching. Three matching methods---covariate-only, single-penalty, and dual-penalty matching---are compared.}
 \label{fig:matching-diag}
\end{figure}

The dual-penalty matching is implemented using the algorithm in~\eqref{eq:opt1} with $\kappa=2$. 
Constraints for matched-set sizes are specified as $(L_\mathcal{C},U_\mathcal{C}) = (1,\infty)$ and $ (L_\mathcal{T},U_\mathcal{T}) = (1,\infty)$. For comparison, we also consider covariate-only matching with $\delta_{ij}$ and single-penalty matching with $D^{(\kappa)}_{ij}$. We evaluate the performance of each matching method by the following metrics: 
(1) covariate dissimilarity, measured by average $\delta_{ij}$ across matched treated-control pairs; and (2) network distance measured by average $d_{ij}$ within matched sets. Figure~\ref{fig:matching-diag} shows the two metrics under the null setting ($\beta_z = 0$) when strong network dependence is present. 

All three matching methods of covariate-only, single-penalty, and dual-penalty, substantially reduce covariate dissimilarity compared with no matching. The dual-penalty matching yields slightly greater covariate dissimilarity than the other two methods because additional penalties on network distance reduce the number of feasible matches.
In return, it increases the network distance within matched sets. These results illustrate the key design trade-off: a modest loss in covariate similarity can produce a substantial reduction in within-matched-set network dependence. Results for the other two dependence levels are presented in Supplementary Material~\ref{apdx::sim:setting}.

\subsection{Randomization-based inference results} \label{sec:sim_results2}

For each matched design, we test the null hypothesis in~\eqref{eq:nullM} using randomization inference with covariance adjustments parameterized by $(\eta,\rho)$ as specified in~\eqref{eq:modelboundM}. We also consider the unadjusted case, in which all covariances are set to zero.  
For comparison, we fit an unmatched linear regression of $\bY$ on $\bZ$ and, for each matched design, a regression that additionally includes a matched-set indicator. We also perform oracle randomization inference based on the known covariance matrix $\bV$, using Monte Carlo sampling to approximate the randomization distribution of $T(\bZ,\bY)$; see Supplementary Material~\ref{apdx:permutation} for details. This procedure provides a benchmark for evaluating the performance of the matching methods when the randomization distribution of the test statistics is approximated using the true covariance structure of treatment assignments.
The performance of each inferential procedure is evaluated using its empirical rejection rates, defined as the proportion of $p$-values below 0.05 across 500 replications.

\begin{table}[!ht]
    \centering
    \caption{\small Empirical rejection rates under the sharp null hypothesis of no causal effect for an unmatched linear regression and three inferential procedures applied after each of three matching designs, across levels of network dependence.}
    \label{tab:NULL}
  \resizebox{0.9\textwidth}{!}{
  \begin{tabular}{ll|ccc}
\toprule
\textbf{Matching design}
& \textbf{Inference procedure}
& \multicolumn{3}{c}{\textbf{Network-dependence level}} \\
\cmidrule(lr){3-5}
&
& \textbf{Independent}
& \textbf{Moderate}
& \textbf{Strong} \\
\midrule
& Unmatched linear regression & 0.064 & 0.140 & 0.242\\\hline \hline
\multirow{3}{*}{Covariate-only}
 & Oracle randomization inference &0.046 & 0.066& 0.078 \\ 
  & Linear regression with matched-set indicators  & 0.048 & 0.110 & 0.190 \\
  & Unadjusted randomization inference         & 0.050 & 0.092 & 0.180 \\
\midrule

\multirow{3}{*}{Single-penalty}
& Oracle randomization inference  &0.056&0.062& 0.076\\ 
  & Linear regression with matched-set indicators & 0.052 & 0.106 & 0.196 \\
  & Unadjusted randomization inference             & 0.054 & 0.104 & 0.174 \\
  
\midrule

\multirow{3}{*}{Dual-penalty}
 & Oracle randomization inference &0.052&0.054 & 0.058\\ 
  & Linear regression with matched-set indicators  & 0.062 & 0.122 & 0.214 \\
  & Unadjusted randomization inference              & 0.052 & 0.106 & 0.196 \\
\bottomrule
\end{tabular}}
\end{table}

Table~\ref{tab:NULL} presents the rejection rates across three levels of network dependence. The unmatched linear regression exhibits substantial Type I error inflation under network dependence, with the degree of inflation increasing as network dependence strengthens. Matching generally mitigates this inflation; however, linear regression with matched-set indicators and unadjusted randomization inference fail to control the Type I error across all three matching methods. 
Under the oracle randomization inference, covariate-only and single-penalty matching produce slightly inflated rejection rates, suggesting that close network connections between matched units may induce spurious association within matched-set comparisons, which persist despite correct adjustment for dependence across matched sets. On the other hand, dual-penalty matching yields rejection rates closest to the nominal level across the three dependence levels under the oracle randomization inference. These findings demonstrate the effectiveness of the proposed dual-penalty matching method in reducing within-matched-set network dependence and highlight the need to adjust for covariance across matched sets to achieve valid inference.  

\begin{figure}[!ht]
    \centering
\includegraphics[width=1.0\linewidth,height=0.28\textheight,keepaspectratio]{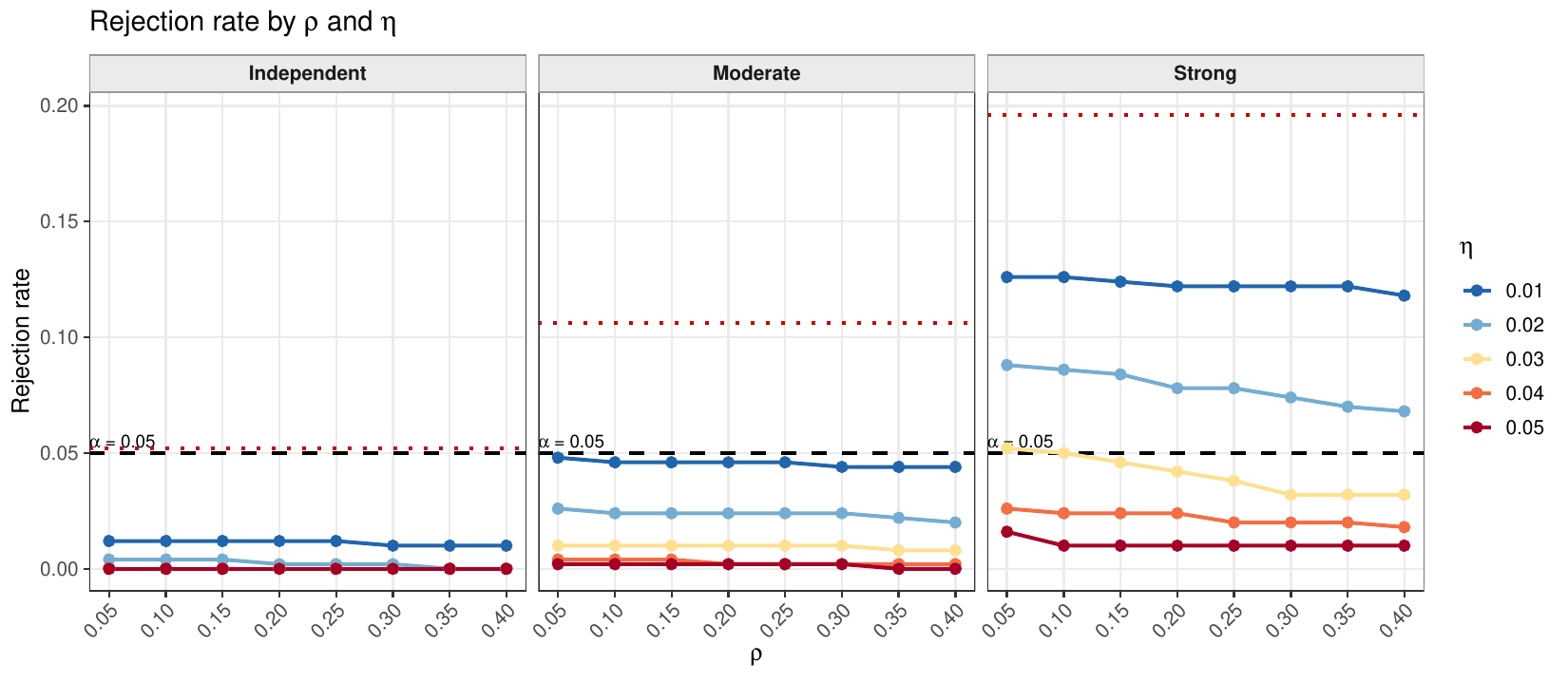}
    \caption{Rejection rates for randomization inference with covariance adjustments parameterized by $(\eta,\rho)$ across levels of network dependence. The dashed line marks the nominal significance level $\alpha=0.05$, and the dotted line represents the rejection rate from unadjusted randomization inference.}
    \label{fig:counterplot}
\end{figure}

Figure~\ref{fig:counterplot} presents the rejection rates from randomization inference with dual-penalty matching over a grid of parameter values, with $\eta \in [0.01, 0.05]$ and $\rho \in [0.05, 0.40]$.
The black dashed lines indicate the nominal significance level, $\alpha=0.05$, and the red dotted lines indicate the rejection rate from unadjusted randomization inference. Under no network dependence (\emph{Independent}) and moderate network dependence, dependence-adjusted randomization inference produces conservative $p$-values across the considered range of $(\eta, \rho)$, with rejection rates below $\alpha$ = 0.05.   
Under strong dependence, however, small values of $\eta$ (e.g., $\eta \leq 0.02$) produce inflated Type I error rates across the considered range of $\rho$, suggesting that the covariances across matched sets are underestimated. Across all levels of network dependence, $\eta$ has a greater effect on the rejection rates than $\rho$. This is because $\eta$ controls the overall magnitude of the covariance bound, whereas $\rho$ controls its decay with network distance. Overall, the figure suggests that underestimating the covariances, for example, by choosing a small value of $\eta$, may lead to inflated Type I errors, whereas large values generally result in conservative inference.

\section{Data application}\label{sec:application}
\subsection{Framingham Heart Study and analysis cohort}

Beginning with the enrollment of its original cohort in 1948, the Framingham Heart Study (FHS) is an ongoing longitudinal study originally designed to investigate cardiovascular epidemiology~\citep{mahmood2014framingham}. It comprises more than 15,000 participants across multiple generations from the town of Framingham, Massachusetts, and neighboring communities. Study participants are therefore likely to be closely connected within these communities. Participants are followed through repeated examinations conducted at intervals of approximately two to eight years.
In addition to its contributions to cardiovascular epidemiology, the FHS has played an important role in ``network'' research, including studies of social contagion, using the information on social ties available in the data that enable the reconstruction of the partial social networks among study participants~\citep{christakis2007spread,christakis2008collective,fowler2008dynamic, o2014estimating}. Existing studies on social contagion or peer effects---which refers to the causal effect of one individual's outcome on the outcomes of socially connected individuals---may provide evidence on the presence of network dependence and/or interference among study participants~\citep{lee2021network,christakis2007spread,ogburn2024causal}.

As an illustrative example, we examine the causal effect of obesity on height-indexed left ventricular mass (LVM) by testing the sharp null hypothesis in the study sample.  
We use data from participants in the FHS Offspring Cohort collected during its fourth and fifth examinations, with Exam 4 (1987-1991) serving as the baseline and Exam 5 (1991-1995) as the follow-up.
Obesity status, defined as a body mass index of at least 30 kg/m$^2$, is measured at baseline. The outcome, height-indexed LVM (g/m), is measured at the follow-up examination.
The baseline covariates include age, sex, smoking status, alcohol use, systolic blood pressure and baseline height-indexed LVM.  We conduct a complete-case analysis, retaining only participants with no missing values for the treatment, outcome or baseline covariates. The final analytic cohort contains 2,347 participants, of whom 371 had obesity (treated) and 1,976 did not (control) at baseline. We use all types of network ties recorded in the data to construct a single network, resulting in 3,817 edges among the 2,347 participants.
Supplementary Material~\ref{apdx:fhs:cohort-network} provides further details on the construction of the analytic cohort and summary statistics of key variables and network.

The na\"ive analysis, which matches participants using baseline covariates only and performs standard randomization-based inference, results in a $p$-value of $9.88\times10^{-9}$, leading to the rejection of the null hypothesis of no causal effect. To adjust for potential network dependence in both matching and inference, 
we use dual-penalty matching with a threshold of $\kappa$=3, motivated by the previous work suggesting that peer effects may extend up to a network distance of three~\citep{christakis2007spread, fowler2010cooperative}. We then conduct randomization-based inference with the covariances adjusted for dependence. Using the design-based upper bounds on the covariances, corresponding to $(\eta,\rho) = (1,1)$ in~\eqref{eq:modelboundM}, we obtain an upper-bound $p$-value of $0.625$. This result indicates that accounting for the maximum permitted statistical dependence across matched sets can increase the $p$-value to as much as $0.625$. We then vary $(\eta, \rho)$ with $\eta\in[0.01,0.05]$ and $\rho\in[0,1]$ to examine how the $p$-value changes.

\begin{figure}[!ht]
    \centering
        \centering     \includegraphics[width=\linewidth,height=0.20\textheight,keepaspectratio]{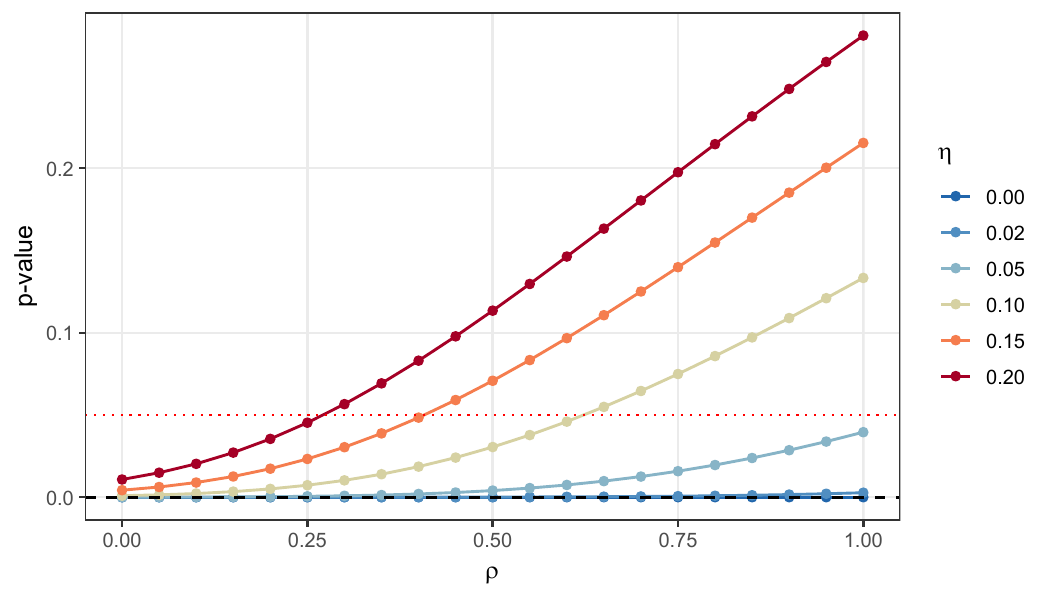}
    \caption{The $p$-value across values of $(\eta,\rho)$ when the covariances among the test statistics are adjusted for network dependence according to these parameters.}
    \label{fig:fhs-sensitivity}
\end{figure}

Figure~\ref{fig:fhs-sensitivity} shows how the $p$-values change when randomization-based inference is performed at varying values of $(\eta, \rho)$. The figure demonstrates that the null hypothesis is rejected for all $\rho \in [0,1]$ when $\eta < 0.05$. 
When $\eta \geq 0.05$, the null hypothesis is more likely not to be rejected as $\rho$ increases—that is, as dependence decays more slowly with network distance.
Supplementary Material~\ref{apdx:fhs:thresholds} provides additional computational details and additional results across alternative thresholds.

\section{Discussion}\label{sec:discussion}

In this work, we provide a unified framework from matching to randomization-based inference for valid causal inference with observational network data. Our methods aim to reduce spurious associations due to network dependence through matching with network-proximity penalties and to address potentially correlated test statistics for valid inference. 
The relevance of the research question depends on the extent to which the key variables share similar autocorrelation structures. The problem of spurious associations becomes particularly evident when covariates, treatments, and outcomes exhibit similar autocorrelation patterns. In the FHS application, for example, suppose that participants connected through network ties have similar baseline covariate values, obesity status and also LVM values, even in the absence of a causal effect of obesity on LVM. Then a na\"ive analysis that ignores this shared autocorrelation may construct matched sets in a way that worsens spurious associations between obesity and LVM. Subsequent post-matching analyses may also underestimate the variances. On the other hand, if one of these variables is largely independent of the underlying network, the problem of spurious association may be less evident.

There are several limitations to this work.
First, our proposed matching method requires selecting a threshold $\kappa$ when defining distance measures used in network proximity penalties.
In practice, when the network is dense, specifying a relatively large $\kappa$ may substantially reduce the number of feasible matches because fewer pairs satisfy the network distance constraints. This reduction in the number of feasible matches may also affect overall covariate similarity after matching. 
One possible extension is to allow different thresholds for between- and within-treatment-arm penalties, particularly when these two sources of dependence contribute differently to spurious treatment--outcome associations.
Moreover, the validity of the proposed randomization-based inference relies on uniform treatment assignment within matched sets. When matching is inexact or unmeasured confounding remains, the assumption of a uniform distribution of treatment assignments may not hold. Extending randomization-based inference to settings with inexact matching under network dependence is an important direction, building on existing work in non-network settings~\citep{pimentel2024covariate,zhu2023randomization,heng2025sensitivity}. 

\section*{Data availability}
The \texttt{R} package \texttt{netmatchRI} is available at \url{https://github.com/zhejiadong/netmatchRI}. 
The FHS data used in this analysis are available to investigators through the dbGaP Authorized Access system under study accessions \href{https://www.ncbi.nlm.nih.gov/projects/gap/cgi-bin/study.cgi?study_id=phs000007.v35.p16}{\texttt{phs000007.v35.p16}} and \href{https://www.ncbi.nlm.nih.gov/projects/gap/cgi-bin/study.cgi?study_id=phs000153.v9.p8}{\texttt{phs000153.v9.p8}}. The FHS data are controlled-access and are not redistributed with this article; access requires authorization through dbGaP.

\section*{Acknowledgments}
The Framingham Heart Study is conducted and supported by the National Heart, Lung, and Blood Institute (NHLBI) in collaboration with Boston University (Contract No. N01-HC-25195, HHSN268201500001I, and 75N92019D00031). This manuscript was not prepared in collaboration with investigators of the Framingham Heart Study and does not necessarily reflect the opinions or views of the Framingham Heart Study, Boston University, or NHLBI.

\putbib
\end{bibunit}

\newpage
\begin{bibunit}
\begin{appendix}

\setcounter{equation}{0}
\setcounter{figure}{0}
\setcounter{table}{0}
\setcounter{page}{1}
\setcounter{section}{0}
\setcounter{theorem}{0}
\setcounter{lemma}{0}
\setcounter{corollary}{0}
\renewcommand{\theequation}{S\arabic{equation}}
\renewcommand{\thefigure}{S\arabic{figure}}
\renewcommand{\thesection}{S\arabic{section}}
\renewcommand{\thetable}{S\arabic{table}}
\renewcommand{\thelemma}{S\arabic{lemma}}
\renewcommand{\thecorollary}{S\arabic{corollary}}
\renewcommand{\thetheorem}{S\arabic{theorem}}
\renewcommand{\theHlemma}{S\arabic{lemma}}
\renewcommand{\theHcorollary}{S\arabic{corollary}}
\renewcommand{\theHtheorem}{S\arabic{theorem}}
\renewcommand{\theHassumption}{S\arabic{assumption}}
\renewcommand{\thepage}{S\arabic{page}}

\begin{center}
{\Large
\textbf{Supplementary Materials}
}
\end{center}

\section{Proofs}
    
\label{apdx:proof}
\subsection{Proof of Corollary~\ref{cor::ignorability}}
Fix a unit $i$ in a matched set $s$ and an assignment vector $\bz'\in \Omega_\mathcal{M}$. 
By Condition~\ref{condition::separation}, every pair of distinct units in matched set $s$ is separated by a network distance greater than $\kappa\ge d_0$. Condition~\ref{assumption::local} therefore implies the treatment assignments of the units within matched set $s$ are jointly independent. Moreover, Assumption~\ref{assumption::interference} implies that
$Y_{si}(\bz')=Y_{si}(\bz'_{\mathcal N_{d_0}(si)}),
$
so the potential outcome of unit $i$ depends on $\bz'$ only through the treatment assignments within its  $d_0$-neighborhood.
Assumption~\ref{assumption::inditreatment} implies, for both $z \in \{0,1\}$, $\Pr(Z_{si} = z \mid \bX_{\mathcal{N}_{d_0}(si)}) = \Pr(Z_{si} = z\mid \bX_{si})$.
Because $Z_{si} \in \bZ_{\mathcal N_{d_0}(si)}$, Assumption~\ref{assumption::ignorability} implies  
$$
\Pr\!\left\{Z_{si}=z\mid Y_{si}(\bz'),\bX_{\mathcal N_{d_0}(si)}\right\}
=
\Pr\!\left\{Z_{si}=z\mid\bX_{\mathcal N_{d_0}(si)}\right\}.
$$
Write $
\bX_{\mathcal N_{d_0}(si)}
=
(\bX_{si},\bX_{\mathcal N_{d_0}(si)\setminus\{si\}})$.  By the law of total probability, for any $\bz'$ and $z$,

\begin{align*}
&\Pr\!\left\{
Z_{si}=z\mid Y_{si}(\bz'),\bX_{si}
\right\} \\
&\quad=
\int
\Pr\!\left\{
Z_{si}=z
\mid
Y_{si}(\bz'),\bX_{si},
\bX_{\mathcal N_{d_0}(si)\setminus\{si\}}=\bx
\right\}
\,d\Pr\!\left\{
\bx\mid Y_{si}(\bz'),\bX_{si}
\right\} \\
&\quad=
\int
\Pr\!\left\{
Z_{si}=z
\mid
\bX_{si},
\bX_{\mathcal N_{d_0}(si)\setminus\{si\}}=\bx
\right\}
\,d\Pr\!\left\{
\bx\mid Y_{si}(\bz'),\bX_{si}
\right\} \\
&\quad=
\int
\Pr\!\left\{
Z_{si}=z\mid\bX_{si}
\right\}
\,d\Pr\!\left\{
\bx\mid Y_{si}(\bz'),\bX_{si}
\right\} \\
&\quad=
\Pr\!\left\{
Z_{si}=z\mid\bX_{si}
\right\}.
\end{align*}
The second equality follows from conditional treatment ignorability given
$\bX_{\mathcal N_{d_0}(si)}$, whereas the third follows from the reducible
treatment-propensity assumption.
Hence, $Y_{si}(\bz')\perp\!\!\!\perp Z_{si}\mid\bX_{si}$ as stated. \hfill$\square$
\medskip

\subsection{Proof of Lemma~\ref{lemma::moments}}\label{proof::lemma}
Fix a matched set $s$. Under Conditions~\ref{assumption::local} and ~\ref{condition::separation}, the treatment assignments of the units
within matched set $s$ are jointly independent. Condition~\ref{assumption::exact} further implies that all units in the matched set have a common treatment probability $p_s$. Under the matched design, the value of $p_s$ is fixed at $n_{st}/n_s$. Therefore, $\bZ_s$ follows a uniform distribution over $\Omega_s$,
$$\Pr(\bZ_s) = \binom{n_s}{n_{st}}^{-1}.$$

We next derive the randomization distribution of $U_s$.
Under $H_{0,\mathcal{M}}$, all potential outcomes are fixed, $Y_{si}(\bZ) = Y_{si}$. Therefore, $f_{sij\bZ} = \mathbf{1}\{Y_{si} > Y_{sj}\}$ is fixed. Assume there are no tied outcomes within matched sets. Let $R_{si} = 1 + \sum^{n_s}_{j=1}f_{sij}$ be the outcome rank of unit $i$ in matched set $s$. Because $\bZ_s$ is uniform over $\Omega_s$, the ranks corresponding to the $n_{st}$ treated units form a simple random sample without replacement from $\{1,\ldots,n_s\}$.
 Let $W_{s}= \sum_{i=1}^{n_s}Z_{si}R_{si}$. Then $U_s =  W_s - \frac{n_{st}(n_{st}+1)}{2}$.
Therefore, $U_s = k$ if and only if $W_s = k + \binom{n_{st}+1}{2}$, i.e., the treated ranks sum to $k + \binom{n_{st}+1}{2}$. 
By definition, $c(n_{st},n_{sc},k)$ counts the number of $n_{st}$-element subsets of $\{1,\ldots,n_s\}$ with this sum, giving
$$\Pr(U_s = k) = \frac{c(n_{st}, n_{sc}, k)}{\binom{n_{st}+n_{sc}}{n_{st}}}, \quad k = 0, 1, \ldots, n_{st}n_{sc}.$$
Each rank $r \in \{1,\ldots,n_s\}$ is equally likely to appear in the treated group. Therefore, the rank of each treated unit has a marginal mean of $(n_s+1)/2$, and hence
\begin{align*}
\E[W_s] = n_{st} \cdot \frac{n_s+1}{2}
\end{align*}
Consequently,
\begin{align*}
    \E[U_s] &= n_{st}\cdot\frac{n_s+1}{2} - \frac{n_{st}(n_{st}+1)}{2} \\& = \frac{n_{st}n_{sc}}{2}.
\end{align*}
Since $U_s = W_s - \frac{n_{st}(n_{st}+1)}{2}$, $\V[U_s] = \V[W_s]$. Each $Z_{si}$ follows a Bernoulli$(n_{st}/n_s)$ distribution. Then,
\begin{align*}
    \E[Z_{si}] = \frac{n_{st}}{n_s}, \quad
\V[Z_{si}] = \frac{n_{st}n_{sc}}{n_s^2}, \quad
\end{align*}
and
\begin{align*}
\mathrm{Cov}(Z_{si},Z_{sj}) &= \E[Z_{si}Z_{sj}] - \E[Z_{si}]\E[Z_{sj}] \\
& = \frac{n_{st}(n_{st}-1)}{n_s(n_s-1)} - \frac{n_{st}^2}{n_s^2} \\&= -\frac{n_{st}n_{sc}}{n_s^2(n_s-1)}, \quad i\neq j.
\end{align*}
Expanding the variance of $W_s = \sum_{i=1}^{n_s} Z_{si}R_{si}$,
\begin{align*}
\V[W_s]
&= \sum_{i=1}^{n_s} R_{si}^2\,\V[Z_{si}] + \sum_{\substack{1\le i,j\le n_s\\i\ne j}} R_{si} R_{sj}\,\mathrm{Cov}(Z_{si},Z_{sj})
\\& = \frac{n_{st}n_{sc}}{n_s^2}\sum^{n_s}_{i}R_{si}^2 - \frac{n_{st}n_{sc}}{n_s^2(n_s-1)}\sum_{\substack{1\le i,j\le n_s\\i\ne j}}R_{si}R_{sj}.
\end{align*}
Using the fact that $\sum_{i\neq j}R_{si}R_{sj} = \bigl(\sum_i R_{si}\bigr)^2 - \sum_i R_{si}^2$, we have
\begin{align*}
\V[W_s]
= \frac{n_{st}n_{sc}}{n_s^2(n_s-1)}\Bigl[n_s\sum^{n_s}_{i}R_{si}^2 - \Bigl(\sum^{n_s}_i R_{si}\Bigr)^2\Bigr].
\end{align*}
Since $R_{si} \in \{1,\ldots,n_s\}$, we have $\sum_i R_{si} = \tfrac{n_s(n_s+1)}{2}$ and $\sum_i R_{si}^2 = \tfrac{n_s(n_s+1)(2n_s+1)}{6}$. 
Therefore,
\begin{align*}
n_s\sum_i R_{si}^2 - \Bigl(\sum_i R_{si}\Bigr)^2
&= \frac{n_s^2(n_s+1)(2n_s+1)}{6} - \frac{n_s^2(n_s+1)^2}{4} \\
&= n_s^2(n_s+1)\left[\frac{2(2n_s+1)}{12} - \frac{3(n_s+1)}{12}\right]\\
&= \frac{n_s^2(n_s+1)(n_s-1)}{12}.
\end{align*}
Consequently,
\begin{align*}
\V[W_s] &= \frac{n_{st}n_{sc}}{n_s^2(n_s-1)}\cdot\frac{n_s^2(n_s+1)(n_s-1)}{12} \\
&= \frac{n_{st}n_{sc}(n_s+1)}{12}. 
\end{align*}
Since $U_s$ differs from $W_s$ by a constant,
$$\V[U_s] = \frac{n_{st}n_{sc}(n_s+1)}{12}. $$
\hfill$\square$
\medskip

\subsection{Additional regularity conditions and proof of Theorem~\ref{theorem::clt}}\label{apdx:cltproof}

For each matched set $s$, define 
$\mathcal{I}_d(s) = \{l: \min \{ d_{ij}, i \in s ,j \in l \}\leq d\}$
as the $d$-neighborhood of matched set $s$. If $l \notin \mathcal{I}_{d_0}(s)$
Conditions~\ref{assumption::local} and~\ref{condition::separation} imply $\bZ_s \bigCI \bZ_l$ and hence $U_s\perp\!\!\!\perp U_l$, so that $\sigma_{sl}=0$. This result is straightforward if we let $I_1$ and $I_2$ in Condition~\ref{assumption::local} collect all unit indices from the matched sets $s$ and $l$, respectively.
Because $\kappa\ge d_0$, we have $\mathcal{I}_{d_0}(s) \subseteq \mathcal{I}_{\kappa}(s)$, which implies $\sigma_{sl} =0$ whenever $l\notin\mathcal{I}_{\kappa}(s)$.
Thus, for each matched set $s$, $\mathcal{I}_{\kappa}(s)$ contains every matched set whose statistic can have nonzero covariance with $U_s$. Substituting 
this result into~\eqref{eq:EVofT} yields 

\begin{equation*}
\sigma^2_T = \sum_{s=1}^S w_s^2\sigma_{ss} + \sum_{\substack{s<l\\ l\in\mathcal{I}_{\kappa}(s)}} 2 w_s w_l\,\sigma_{sl}.
\end{equation*}

\begin{condition}[Uniformly bounded matched-set sizes]\label{condition::boundedmatchedsets}
There exists a constant $n_{\max}<\infty$, independent of $n$ and $S$, such that \[\max_{1\le s\le S} n_s \le n_{\max}.\]
\end{condition}
\noindent 
\noindent This uniform bound prevents a single matched set from dominating the mean and variance of the statistic. It holds directly when the matched design has bounded set sizes. 
\begin{condition}[Bounded matched-set neighborhood]\label{condition::boundedaffinity}
There exists a constant $\bar\Delta<\infty$, independent of $n$ and $S$, such that $$\max_{1\le s\le S}|\mathcal{I}_{\kappa}(s)|\le\bar\Delta$$
\end{condition}
\noindent 
This condition bounds the size of $\mathcal{I}_\kappa(s)$. It holds when the observed network has bounded degree and the matched-set sizes are bounded, as in Condition ~\ref{condition::boundedmatchedsets}. 

\begin{condition}[Non-degenerate variance]\label{condition::nondevaraince}
There exists a constant $c>0$, independent of $n$ and $S$, such that
\[
\liminf_{S\to\infty}\frac{\sigma^2_T}{S}\ge c
\]
\end{condition}
\noindent Condition~\ref{condition::nondevaraince} rules out cases in which the statistic $T(\bZ,\bY)$ stops accumulating variation as $S$ grows, ensuring that $\sigma^2_T$ grows at least on the order of \(S\). A similar condition for non-degenerate variance is assumed in~\cite{aronow2017estimating}. A sufficient condition for Condition~\ref{condition::nondevaraince} is $\sigma_{sl}\geq 0$ for all $s \neq l$. Such nonnegative dependence is typically assumed in observational network studies~\citep{ogburn2018challenges,lee2021network}. 

We establish asymptotic normality following~\cite {chandrasekhar2023general}. For simplicity, we refer to~\cite{chandrasekhar2023general} as CJMT. Specifically, we prove the four conditions required to apply Theorem 1 from CJMT. With a slight abuse of notation, define $W_s = w_s(U_s-\mu_s)$. 
\paragraph{Step 1: Bounded moment ratio}
We start by showing the moment ratio $\E[|W_{s}|^3]/\{\E [W^2_{s}]\}^{3/2}$ is bounded above.

By definition, $|W_s| = w_s|U_s-\mu_s|\leq w_s\cdot\frac{n_{st}n_{sc}}{2}$. Condition~\ref{condition::boundedmatchedsets} further implies
\begin{align*}
    |W_{s}|  &\leq w_s\cdot\frac{n_{st}n_{sc}}{2}\\& \leq  \frac{n^2_{s}/4}{2(n_{s} + 1)} 
    \\&\leq \frac{1}{8} n_{\max} =: B < \infty.
\end{align*}
Thus, $|W_s|$ is uniformly bounded by $B$ for every $s$.
Moreover,
\begin{align*}
     \dfrac{\E[|W_{s}|^3]}{\{\E [W^2_{s}]\}^{3/2}} & = \dfrac{\E [ |W_{s}| \cdot W^2_{s}]}{\{\E [W^2_{s}]\}^{3/2}} \\&\leq \dfrac{B\, \E W^2_{s}}{\{\E [W^2_{s}]\}^{3/2}} \\&= \frac{B}{\{\E [W^2_{s}]\}^{1/2}}.
\end{align*}
By Lemma~\ref{lemma::moments}, $\E[W_s] = 0$ and $\E[W_s^2] = w^2_s\sigma_{ss}$. Hence,
$$\{\E [W^2_{s}]\}^{1/2} = \{\dfrac{n_{st} n_{sc}}{12(n_{s} + 1)}\}^{1/2 }\ge \{12(n_{\max}+1)\}^{-1/2},$$
where the inequality holds as each matched set contains at least one treated and one control unit, i.e., $n_{st}n_{sc}\ge 1$, and $n_s\leq n_{\max}$ by Condition~\ref{condition::boundedmatchedsets}.
Therefore, the moment ratio required by CJMT is bounded above:
\[\dfrac{\E|W_{s}|^3}{\{\E [W^2_{s}]\}^{3/2}} \le B\,\sqrt{12(n_{\max} + 1) } =: B_1 <\infty \]

\paragraph{Step 2: Assumption~1 in CJMT.}
We aim to show
\[\Sigma_1  := \sum_{s=1}^{S} \sum_{j,l\in\mathcal{I}_{\kappa}(s)}
\mathbb{E}\!\left[|W_{s}|W_{j}W_{l}\right] =   o((\sigma_{T}^2)^{3/2})
\]
It suffices to show that 
$$\frac{|\Sigma_1|}{(\sigma^2_T)^{3/2}}\to 0.$$
Recall from Step 1 that $|W_s|$ is uniformly bounded by the constant $B$. 
Therefore, by the triangle inequality
\begin{align*}
        |\Sigma_1| \leq &\sum^S_{s=1}\sum_{j,l\in\mathcal{I}_{\kappa}(s)}\E\big[|W_s||W_j||W_l|\big]\\
         \leq & B^3\sum^S_{s=1}|\mathcal{I}_{\kappa}(s)|^2.
\end{align*}

By Condition~\ref{condition::boundedaffinity}, $\max_{1\leq s\leq S} |\mathcal{I}_\kappa(s)|\leq \bar \Delta$, and hence
$$ |\Sigma_1| \leq S \bar\Delta^2B^3.$$
By Condition~\ref{condition::nondevaraince}, $(\sigma^2_{T})^{3/2} = \sigma^3_{T}\geq (cS)^{3/2}$.
Consequently,
\[\dfrac{|\Sigma_1|}{(\sigma_{T}^2)^{3/2}} \leq \dfrac{S\bar \Delta^2 \,B^3}{(cS)^{3/2}} = O(S^{-1/2}) \to 0. \]
Therefore,
$$\Sigma_1 = o((\sigma_{T}^2)^{3/2}).$$

\paragraph{ Step 3: Assumption~2 in CJMT.}
We aim to show 
\[\Sigma_2  := \sum_{s,j=1}^{S}
\sum_{\substack{
l \in \mathcal{I}_{\kappa}(s) \\
q \in \mathcal{I}_{\kappa}(j)
}}
\text{Cov}(W_{s}W_{l},W_{j}W_{q}) =   o((\sigma_{T}^2)^2)
\]
Let $I_1$ and $I_2$ in Condition~\ref{assumption::local} collect all indices of matched sets $\{s,l\}$, $\{j,q\}$, respectively. 
Therefore,
$$\text{Cov}(W_sW_l,W_jW_q) = 0$$
unless at least one of $j$ or $q$ belongs to $\mathcal{I}_\kappa(s)\cup \mathcal{I}_\kappa(l)$. We now investigate the nonzero covariance terms. Condition~\ref{condition::boundedaffinity} implies there are at most $\bar\Delta$ choices of $l \in   \mathcal{I}_\kappa(s)$. For fixed $s$ and $l$, there are at most $2\bar\Delta$ choices of $j$ with potential nonzero covariances,
as $|\mathcal{I}_\kappa(s)\cup \mathcal{I}_\kappa(l)|\leq 2 \bar\Delta$. For each such $j$, there are at most $\bar\Delta$ choices of $q\in \mathcal{I}_\kappa(j)$. Hence, the case in which $j$ belongs to $\mathcal{I}_\kappa(s)\cup \mathcal{I}_\kappa(l)$ contributes at most $2\bar\Delta^2$ non-zero covariances.

By symmetry, if $q$ belongs to the union, the number of non-zero covariances is at most $2\bar\Delta^2$.
Therefore, for each fixed pair $(s,l)$, there are at most $4\bar\Delta^2$ nonzero covariances. Because there are at most $S\bar\Delta$ possible pairs $(s,l)$, the total number of potential nonzero covariances in $\Sigma_2$ is bounded by $4S\bar\Delta^3$.

\begin{align*}
      |\text{Cov}(W_{s}W_{l},W_{j}W_{q})| &\leq |\E[W_sW_lW_jW_q]| + |\E[W_sW_l]||\E[W_jW_q]|\\
    & \leq 2B^4.
\end{align*}

Consequently, $|\Sigma_2|\leq 8S\bar\Delta^3B^4$. By Condition~\ref{condition::nondevaraince}, $(\sigma^2_{T})^2 \geq (c S)^2$. Hence,
\[\dfrac{|\Sigma_2|}{(\sigma^2_{T})^2 }\leq \dfrac{8S\bar\Delta^3B^4}{(c S)^2} = O(S^{-1}) \to 0\]

\paragraph{ Step 4: Assumption~3 in CJMT.}
We aim to show 
\[\Sigma_3:= \sum^{S}_{s=1} \E\!\left[\,\left|\mathbb{E}\!\left[W_{s}\sum_{l\notin\mathcal{I}_{\kappa}(s)} W_{l}\,\Big|\,\{W_{l}\}_{l\notin\mathcal{I}_{\kappa}(s)}\right]\right|\,\right] = o(\sigma^2_{T})\]

Let $I_1$ collect all unit indices in the matched set $s$ and $I_2 = \{l: l\notin \mathcal{I}_{\kappa}(s)\}$ in Condition~\ref{assumption::local}, then $W_s$  is  independent of $\{W_{l}\}$ for $l \notin \mathcal{I}_{\kappa}(s)$.
Hence
\[
\mathbb{E}\!\left[W_{s}\sum_{l\notin\mathcal{I}_{\kappa}(s)} W_{l}\,\Big|\,\{W_{l}\}_{l\notin\mathcal{I}_{\kappa}(s)}\right] = \E[W_{s}]\times\sum_{l\notin\mathcal{I}_{\kappa}(s)} W_{l} = 0
\]
Therefore, each summand of $\Sigma_3$ is identically zero. Then $\Sigma_3 =0 $, and trivially $\Sigma_{3} = o(\sigma^2_{T})$. 

\paragraph{Application of Theorem~1 in CJMT}
Step 1 shows that $\E[|W_{s}|^3]/\{\E [W^2_{s}]\}^{3/2}$ is uniformly bounded.
Steps 2-4 establish Assumptions~1,~2, and~3 in CJMT. Moreover, Condition~\ref{condition::nondevaraince} implies  $\sigma^2_T\to \infty$, as $S\to \infty$ (i.e., $n\to \infty$).  Therefore, all conditions of Theorem~1 in CJMT are satisfied. By definition, $T -\mu_T = \sum^S_{s=1 }W_s$, and $\V[\sum^S_{s=1}W_s] = \sigma^2_T$.
Theorem~1 in CJMT yields: 
\[\dfrac{T -\mu_T}{\sigma_T}\xrightarrow{d} \mathcal{N}(0,1)\]
\hfill$\square$
\medskip

\subsection{Proof of Theorem~\ref{theorem::bound}}
\label{ssec:proof2}

Let $F_{s,l}$ denote the joint CDF of $(U_s,U_l)$, with marginal CDFs $F_s$ and $F_l$, respectively, as  determined by Lemma~\ref{lemma::moments}. By the Fréchet–Hoeffding inequality~\citep{frechet1951tableaux}, any bivariate CDF $F_{s,l}$ with fixed marginal distributions satisfies
$$ F_{s,l}(u_s,u_l) \leq \min(F_s(u_s),F_{l}(u_l)) =: F^M_{s,l}(u_s,u_l),$$
where $F^M_{s,l}$ is the joint CDF that maximizes $\text{Cov}(U_s,U_l)$ over all joint CDFs with fixed marginal distributions~\citep{nelsen2006introduction,puccetti2010multivariate}. Write $M_{sl}=\mathrm{Cov}_M(U_s,U_l)$ for the covariance attained under $F^M_{s,l}$ for any $s$ and $l$. 

By Lemma~\ref{lemma::moments}, the marginal CDF $F_s$ of $U_s$ is fully determined by $(n_{st}, n_{sc})$. 
For any joint distribution $F_{s,l}$ of $(U_s, U_l)$ with fixed marginals $F_s$ and $F_l$, the Fr\'{e}chet--Hoeffding inequality gives
$$F_{s,l}(a,b) \leq \min\bigl(F_s(a),\, F_l(b)\bigr) =: F^M_{s,l}(a,b),$$
so $F^M_{s,l}$ is the pointwise largest joint CDF consistent with the fixed marginals~\citep{frechet1951tableaux}. 
Since $\mathrm{Cov}(U_s, U_l) = \E[U_sU_l] - \mu_s\mu_l$,
$$M_{sl} = \E_{M}[U_sU_l] - \mu_s\mu_l$$
where $\E_M[U_sU_l]$ is taken over $F^M_{s,l}$. Because $U_s$ and $U_l$ are discrete,
$$\E_M[U_sU_l] = \sum_{a=0}^{n_{st}n_{sc}}\sum_{b=0}^{n_{lt}n_{lc}} ab\, p^M_{ab},$$
where $p^M_{ab} = \Pr_M(U_s = a, U_l = b)$ and
\begin{align*}
p^M_{ab} &= \Pr_M(U_s = a, U_l = b) \\
&= \Pr_M(U_s \leq a, U_l \leq b) - \Pr_M(U_s \leq a-1, U_l \leq b) \\
&\quad - \Pr_M(U_s \leq a, U_l \leq b-1) + \Pr_M(U_s \leq a-1, U_l \leq b-1)\\
&= F^M_{s,l}(a,b) - F^M_{s,l}(a-1,b) - F^M_{s,l}(a,b-1) + F^M_{s,l}(a-1,b-1),
\end{align*}
with the convention $F^M_{s,l}(-1,\cdot) = F^M_{s,l}(\cdot,-1) = 0$. Therefore,
\[M_{sl} =\sum_{a=0}^{n_{st}n_{sc}}\sum_{b=0}^{n_{lt}n_{lc}} ab\, p^M_{ab} - \mu_s\mu_l\]
which gives~\eqref{eq:designboundM}.

\hfill$\square$
\medskip

\section{Details about proposed methods}

\subsection{Sensitivity analysis on covariance bounds}
\label{ssec:sensitivity}

The parameters $(\eta,\rho)$ can serve as sensitivity parameters governing covariance bounds. 
The proposed sensitivity analysis framework can be summarized in two ways. First, for any prespecified $(\eta,\rho)$, the $p$-value $p_{M^*}(\eta,\rho)$ represents the inferential conclusion obtained under the corresponding bounds $M^*_{sl}(\eta,\rho)$ for all relevant pairs of matched sets $s$ and $l$. Evaluating $p_{M^*}(\eta,\rho)$ over a grid of $(\eta,\rho)$ can directly show how the evidence against the null hypothesis changes with the overall magnitude and decay rate of the cross-matched-set covariances. Second, a practitioner may also want to determine the degree of dependence required to change the inferential conclusion. For a significance level $\alpha$, define the region
$$\mathcal{C}_{\alpha} = \{(\eta,\rho) \in [0,1]^2: p_{M^*}(\eta,\rho) \leq \alpha\},$$
which contains the values of $(\eta,\rho)$ under which the null hypothesis remains rejected. For a fixed $\rho$, solving $p_{M^*}(\eta,\rho) = \alpha $ gives a critical level
\begin{equation*}
\eta^*(\rho)=\frac{\{T(\bZ^{\text{obs}},\bY^{\text{obs}})-\mu_T\}^2/z_{1-\alpha/2}^2\,-\,\sum^S_{s= 1}w_s^2\sigma_{ss}}{\displaystyle\sum_{\substack{s<l\\ l\in\mathcal{I}_{\kappa}(s)}}2w_sw_l\rho^{h_{s,l}-1}M_{sl}}.
\end{equation*}
The curve $\rho\mapsto\eta^*(\rho)$ defines the level-$\alpha$ critical contour. For a given $\rho$, values of $\eta$ below the contour lead to rejection of the null hypothesis, whereas values above the contour do not. Thus, $\eta^*(\rho)$ quantifies the minimum magnitude of across-matched-set covariance, under the decay rate determined by $\rho$, required to overturn the rejection of the null hypothesis.

For each point $(\eta^*(\rho),\rho)$ on the critical contour, the following quantity remains constant:
\begin{equation*}
R_{\mathrm{crit}}
=
\frac{
N_{\kappa}^{-1}\displaystyle\sum_{\substack{s<l\\ l\in\mathcal I_\kappa(s)}} w_sw_lM^*_{sl}(\eta^*(\rho),\rho)
}{
S^{-1}\displaystyle\sum_{s=1}^S w_s^2\sigma_{ss}
},
\end{equation*}
where $N_{\kappa} = \sum_{s<l}\mathbf 1\{l\in\mathcal I_\kappa(s)\}$ denotes the number of relevant matched-set pairs.
We refer to $R_{\mathrm{crit}}$ as the \textit{critical ratio}.
It provides a scalar summary of the amount of across-matched-set covariance required to change the inferential conclusion. If the average weighted covariance bound across relevant matched-set pairs is less than $R_{\mathrm{crit}}$ times the average weighted within-matched-set variance, the null hypothesis remains rejected.
This interpretation assumes that the denominator defining $\eta^*(\rho)$ is positive and that $\eta^*(\rho)\in[0,1]$. Values outside this interval indicate that the inferential conclusion does not change over the specified sensitivity range.

\section{Additional simulation studies}

\subsection{Data-generating process, computation, and additional results}
\label{apdx::sim:setting}

Given the covariance matrix $\bV$ defined in Section~\ref{ssec:simsetting}, we generate an $n\times 3$ covariate matrix $\bX=(X_1,X_2,X_3)$. We add $0.001\bI_n$ to $\bV$ to ensure positive definiteness and rescale it to have a unit diagonal. We retain the notation of $\bV$ for the resulting covariance matrix. The covariate vectors $X_1$ and $X_2$ are independently drawn from $\mathcal{N}(0,\bV)$ and therefore exhibit network dependence, whereas $X_3$ is drawn from $\mathcal{N}(0,\bI_{n})$, where $\bI_n$ denotes the $n\times n$ identity matrix. The length-$n$ treatment assignment vector is
$$\bZ = \mathbf{1}\{\bX\boldsymbol{\gamma}_x + \bepsilon_z \ge c\}, \quad \bepsilon_z \sim \mathcal{N}(0,\bV),$$
where the indicator function is applied elementwise, $\boldsymbol{\gamma}_x = (0.1,0.1,0.1)^T$ represents the effects of the covariates on treatment assignment, and the constant $c$ is chosen such that the treatment prevalence is approximately $0.23$. The length-$n$ outcome vector is then generated as
$$\bY = \bX\boldsymbol{\beta}_x + \bZ\beta_z + \bepsilon_y, \quad \bepsilon_y \sim \mathcal{N}(0,\bV),$$
where $\boldsymbol{\beta}_x=(0.1,0.1,0.1)^T$ specifies the effects of the covariates on the outcome, and $\beta_z$ denotes the true treatment effect. We set $\beta_z=0$ to evaluate Type I error control and vary $\beta_z$ to assess the power of the inferential methods. All random vectors are generated independently unless otherwise specified.
The simulations are conducted on a Windows desktop with an AMD Ryzen 9 7900X processor and 32 GB of RAM. Using Gurobi Optimizer 10.0.1, the MIP for the dual-penalty matching design requires an average of 5.77 seconds to solve.
\begin{figure}[!ht]
    \centering
    \includegraphics[width=\linewidth]{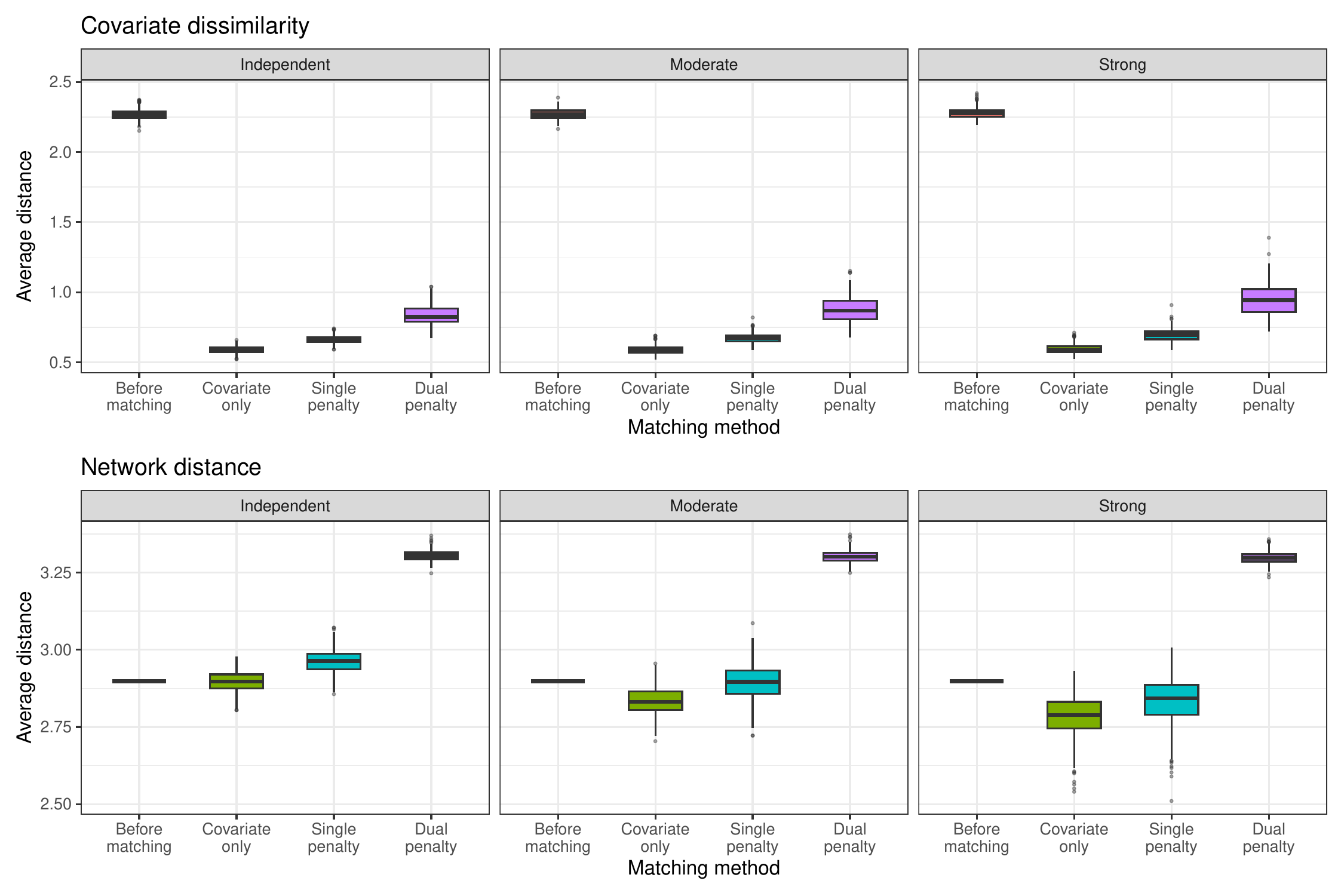}
 \caption{\small Matching diagnostics across three levels of network dependence and under the null ($\beta_z = 0$) over 500 replications.
 }
    \label{fig::apdx:combinned}
\end{figure}

Figure~\ref{fig::apdx:combinned} presents matching diagnostics under the null hypothesis ($\beta_z=0$) across the three levels of network dependence. The results under strong dependence are discussed in Section~\ref{sec:sim_results1}. Similar patterns are observed under independent and moderate dependence. Across all three levels, imposing network-separation constraints reduces the set of feasible matches and therefore produces a modest increase in covariate dissimilarity relative to covariate-only and single-penalty matching.

\subsection{Power analysis under the alternative}
\label{apdx::sim:power}

We evaluate rejection rates under the alternative for four inferential procedures under the dual-penalty 
matching with $\kappa =2$: unmatched linear regression, linear regression with matched-set indicators, unadjusted randomization inference, and adjusted randomization inference with specified $(\eta,\rho)$.  We vary $\beta_z \in \{0, 0.1, 0.2, 0.3, 0.4\}$ over 500 replications under each dependence level. 
We report the results of adjusted randomization inference with $(\eta,\rho) = (0,0)$ under independence (\emph{independent}), $(0.01,0.05)$ under moderate dependence, and $(0.03, 0.10)$ under strong dependence.

\begin{figure}[!ht]
    \centering
    \includegraphics[width=\linewidth]{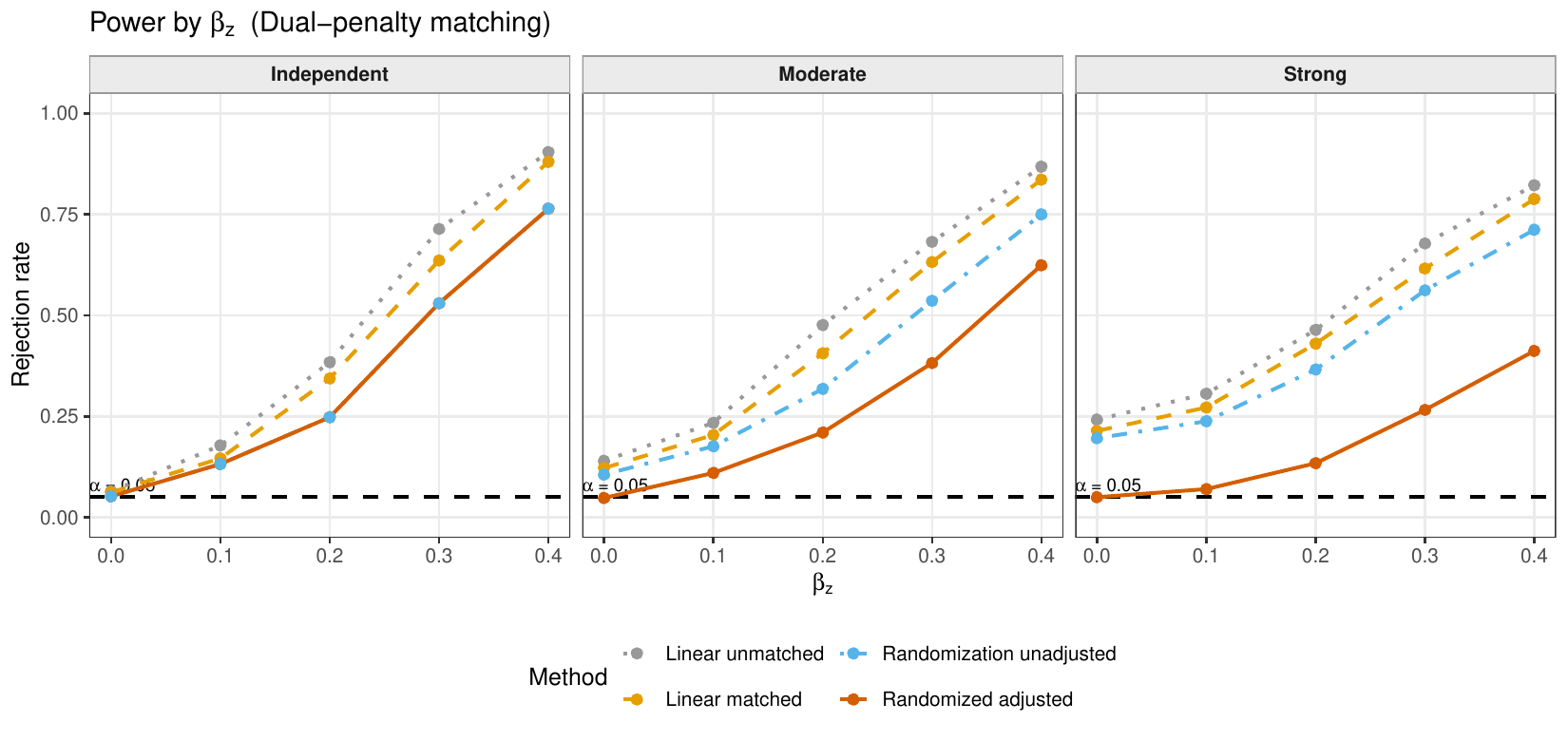}
    \caption{\small Rejection rates as a function of the true treatment effect 
    $\beta_z \in \{0, 0.1, 0.2, 0.3, 0.4\}$ under dual-penalty matching, across 
    three network dependence levels and 500 
    simulation replications. Each line corresponds to one inference method. The dark dashed horizontal line marks the nominal level $\alpha = 0.05$.}
    \label{fig:power}
\end{figure}

Figure~\ref{fig:power} reports the rejection rates as a function of $\beta_z$ for 
each procedure and dependence level. Under all three levels, rejection rates 
increase with $\beta_z$, indicating that all methods have power against 
alternatives away from the null. Under independence, unadjusted and adjusted randomization inference 
coincide because $(\eta,\rho) = (0,0)$. Under moderate and 
strong dependence, adjusted randomization inference with the reported $(\eta,\rho)$ shows lower power than unadjusted randomization inference and the two linear regression models.

\section{Monte Carlo randomization and oracle procedure}~\label{apdx:permutation}

Here, we describe a Monte Carlo approach for approximating the randomization distribution of $T(\bZ,\bY)$ under the null hypothesis. This approach corresponds to the oracle randomization inference procedure described in Section~\ref{sec:simulation}. Unlike the normal-approximation procedure introduced in Section~\ref{sec:inference}, this method requires a fully specified model for the joint distribution of the treatment assignments. Specifically, treatment assignment vectors are generated from a model representing the assumed network dependence structure, and the resulting empirical distribution of the test statistic is used for inference.

We assume that the dependence structure of the treatment assignment can be represented through a latent Gaussian model at the unit level. Let
\[
\bG=(G_1,\ldots,G_n)^\top \sim N(\mathbf 0,\bV_G),
\]
where $\bV_G$ is an $n\times n$ variance-covariance matrix. 
In the oracle procedure, we set $\bV_G$ equal to the true $\bV$ used in generating $\bZ$. In real applications, $\bV_G$ should be specified by the practitioner.
Given the latent variable $\bG$, treatment assignments are generated within each matched set in a manner consistent with the matched design. A natural construction is to assign treatment within each matched set according to the ordering of the latent variables. For a matched set $s$ containing 
$n_{st}$ treated units, the $n_{st}$ units with the largest latent values are assigned treatment, and the remaining $n_{sc}$ units are assigned control. The generated assignment vector $\bZ$ satisfies the matched design by construction, while dependence across matched sets is induced through $\bV_G$. Inference can then be made by comparing the observed test statistic with its empirical distribution based on the simulated $\bZ$.
Algorithm~\ref{alg:mc_gaussian_copula} summarizes the procedure.

\begin{algorithm}[!ht]
\caption{Monte Carlo sampling under a working Gaussian model}
\label{alg:mc_gaussian_copula}
\begin{algorithmic}[1]
\Require Matched sets \(s=1,\ldots,S\), observed outcomes \(\bY^{\mathrm{obs}}\), number of Monte Carlo draws \(B\), working covariance matrix $\bV_G$
\For{\(b=1,\ldots,B\)}
    \State Draw \(\bG^{(b)}=(G_1^{(b)},\ldots,G_n^{(b)})^\top \sim N(\mathbf 0,\bV_G)\)
    \For{each matched set \(s\)}
        \State Let $\bI_s$ denote the set of unit indices in the matched set $s$
        \State Extract the latent values $\{G^{(b)}_i: i\in \bI_s\}$
        \State Assign treatment to the $n_{st}$ units in set $s$ with the largest values of $G^{(b)}_i$
        \State Assign control to the remaining $n_{sc}$ units in set $s$
        \State Denote the resulting treatment assignment vector in set $s$ by $\bZ^{(b)}_s$
    \EndFor
    \State Form \(\bZ^{(b)}=(\bZ_1^{(b)},\ldots,\bZ_S^{(b)})\)
    \State Compute \(T^{(b)} = T(\bZ^{(b)}, \bY^{\mathrm{obs}})\)
\EndFor
\State Compute
\[
p^{(B)}=\frac{1+\sum_{b=1}^B\mathbf{1}\{|T^{(b)}-\mu_T|\ge |T^{\mathrm{obs}}-\mu_T|\}}{1+B}.
\]
\end{algorithmic}
\end{algorithm}

When the latent Gaussian assignment model and $\bV_G$ are correctly specified, this approach may closely approximate the randomization distribution of $T(\bZ,\bY)$ under the matched design. We thus use the Oracle procedure as a benchmark. In practice, however, both the assignment mechanism and $\bV_G$ are unknown, so the validity of the resulting inference is highly sensitive to the misspecification of either component.

\section{Additional details of the FHS data application}\label{apdx:fhs}

\subsection{Descriptive statistics and network summaries}
\label{apdx:fhs:cohort-network}

We required participants in the analytic cohort to have been observed at both Offspring Exams 4 and 5. Of the 3,894 participants who attended Exam 4, 3,420 had follow-up records from Exam 5. We sequentially excluded 17, 562, and 474 participants with missing measurements required to define baseline obesity status, baseline LVM, and follow-up LVM, respectively. We further excluded 20 participants with missing baseline covariates. All remaining participants were at least 21 years old at baseline. The final analytic cohort comprised 2,347 participants, including 371 with obesity and 1,976 without obesity. This complete-case analysis assumes that data on the treatment, outcome, and covariates are missing completely at random.

Table~\ref{tab:fhs-characteristics} presents the distributions of the outcome and baseline covariates stratified by obesity status. Continuous variables are reported as means (standard deviations), and binary variables are reported as counts (percentages). Compared with participants without obesity, those with obesity have a higher mean follow-up height-indexed LVM, a higher mean baseline systolic blood pressure, and a lower proportion of female participants.

\begin{table}[H]
\centering
\normalsize
\begin{tabular}{lrrr}
\toprule
Characteristic & Overall & No obesity & Obesity \\
& $(n=2{,}347)$ & $(n=1{,}976)$ & $(n=371)$ \\
\midrule
Female sex, $n$ (\%) & 1,332 (56.8) & 1,145 (57.9) & 187 (50.4) \\
Past-year regular smoking, $n$ (\%) & 500 (21.3) & 427 (21.6) & 73 (19.7) \\
Any current alcohol use, $n$ (\%) & 1,644 (70.0) & 1,403 (71.0) & 241 (65.0) \\
Systolic blood pressure, mm Hg (SD) & 124.41 (17.86) & 123.03 (17.76) & 131.74 (16.56) \\
Baseline height-indexed LVM, g/m (SD) & 95.28 (20.30) & 93.09 (19.59) & 106.97 (20.06) \\
Follow-up height-indexed LVM, g/m (SD) & 94.49 (21.43) & 92.38 (20.62) & 105.74 (22.18) \\
\bottomrule
\end{tabular}
\caption{Distribution of the outcome and baseline covariates of the analytic cohort stratified by obesity status}
\label{tab:fhs-characteristics}
\end{table}

Table~\ref{tab:fhs-network-summary} summarizes the structural characteristics of the analytic network. The network is sparse, with a mean degree of 3.25, and 14.9\% of participants have no observed network ties.

\begin{table}[H]
\centering
\begin{tabular}{lr}
\toprule
Network statistic & Value \\
\midrule
Participants (nodes) & 2,347 \\
Undirected ties (edges) & 3,817 \\
Degree, mean  & 3.25  \\
Degree, median [range] & 2 [0, 22] \\
Isolated participants, $n$ (\%) & 350 (14.9) \\
Connected components & 500 \\
Largest component, $n$ (\%) & 1595 (68.0) \\
Distance among connected pairs, mean (median) & 7.12 (7) \\
Distance among connected pairs, range & 1--23 \\
\bottomrule
\end{tabular}
\caption{Summary of the analytic cohort network}
\label{tab:fhs-network-summary}
\end{table}

\subsection{Computational implementation and inferential results}\label{apdx:fhs:thresholds}

Dual-penalty matching is conducted with $(L_\mathcal{C},U_\mathcal{C}) = (1,100)$, $(L_\mathcal{T},U_\mathcal{T}) = (1,100)$, and the dissimilarity measure $D^{(3)}_{tc}$. For comparison, we apply covariate-only matching using $\delta_{tc}$ under the same matched-set-size constraints. Both matching designs retain all 2,347 participants and produce 371 matched sets.
The MIP formulation for dual-penalty matching with $\kappa = 3$ involves 720,653 binary variables and 13,463,401 constraints, comprising 11,319,064 control--control constraints, 2,139,643 treated--treated constraints, and 4,694 matched-set size constraints. We solve the MIP using the \texttt{netmatch()} function in the \texttt{netmatchRI} package with the HiGHS solver (version 1.14.0.2) on a Linux workstation (Intel Xeon Platinum 8468 processor and 32 GB RAM). The optimal solution is obtained in 731.5 seconds (12.1 minutes). With the same matched-set size constraints, the corresponding solver times are 128.1 seconds (2.1 minutes) for $\kappa =2$ and 11816.3 seconds (3.3 hours) for $\kappa = 4$. 

\begin{figure}[!h]
    \centering
    \begin{subfigure}[t]{0.49\textwidth}
        \centering
        \includegraphics[width=\linewidth,height=0.23\textheight,keepaspectratio]{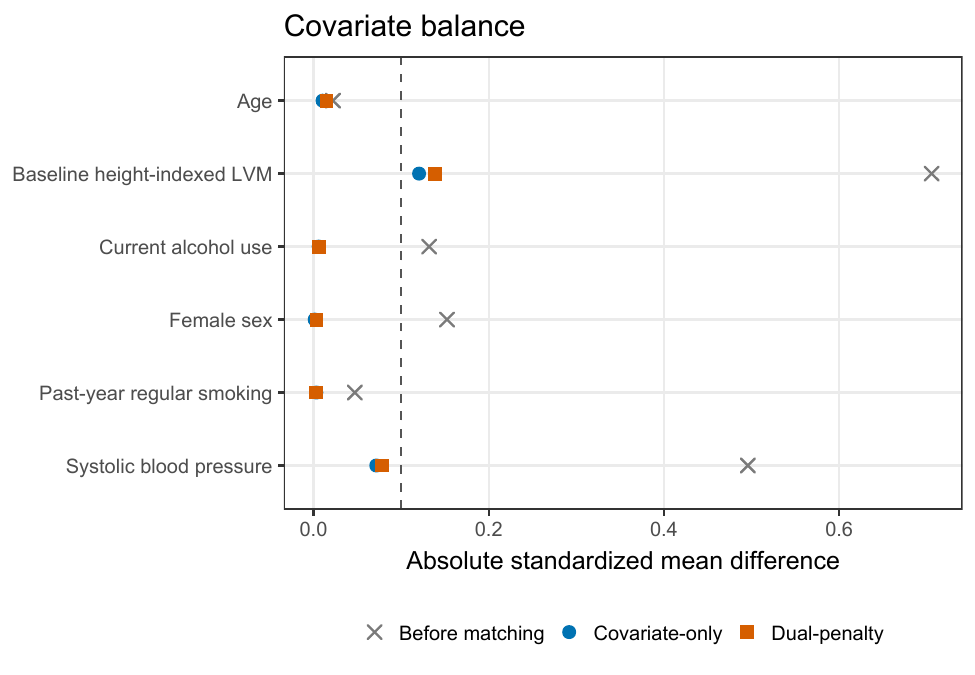}
        \caption{Covariate balance}
        \label{fig:fhs-covariate-balance}
    \end{subfigure}
    \hfill
    \begin{subfigure}[t]{0.49\textwidth}
        \centering\includegraphics[width=\linewidth,height=0.23\textheight,keepaspectratio]{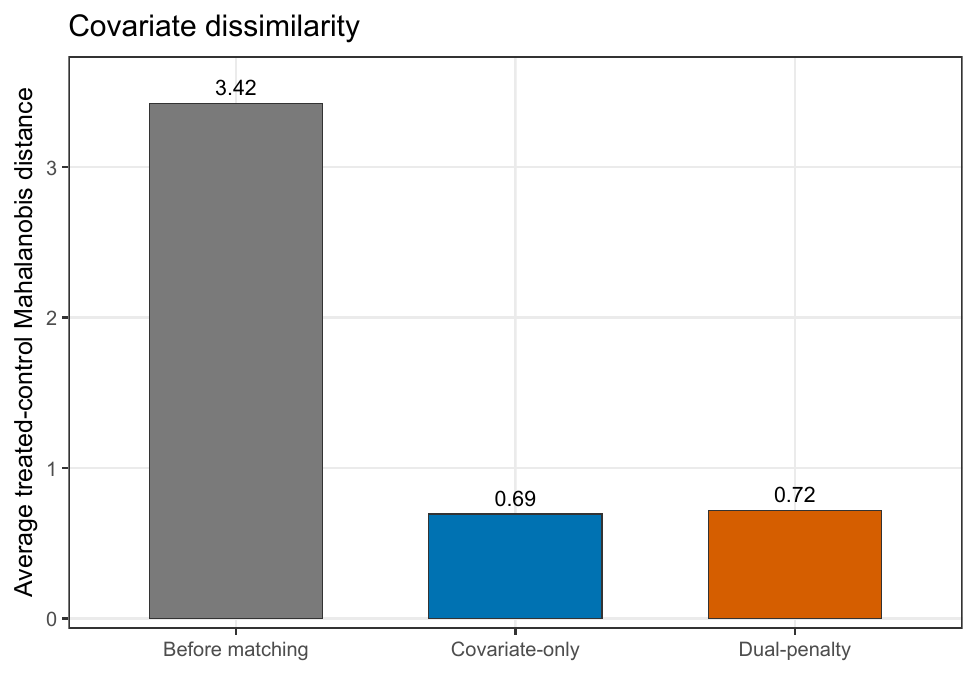}
        \caption{Covariate dissimilarity}
        \label{fig:fhs-covariate-similarity}
    \end{subfigure}
    \caption{Covariate balance and similarity before and after matching in the FHS application. Panel (a) reports the absolute standardized mean difference for each matching covariate; the dashed vertical line marks 0.10. Panel (b) reports the average treated--control Mahalanobis distance.}
    \label{fig:fhs-matching-diagnostics}
\end{figure}

Figure~\ref{fig:fhs-matching-diagnostics} shows covariate balance and covariate similarity before and after matching when applying dual-penalty matching with a threshold of $\kappa$=3.  Compared with covariate-only matching, dual-penalty matching achieves comparable balance and similarity. 

We repeat the primary analysis with $\kappa=3$ using $\kappa=2$ and $\kappa =4$, while retaining the same analytic cohort, variables, and matched-set-size constraints described in Section~\ref{sec:application}. Each dual-penalty matching design retains all 2,347 participants in 371 matched sets. 
Table~\ref{tab:fhs-threshold-inference} presents details of the computational implementation and inferential results for the three threshold values. All MIP problems are solved using HiGHS. The na\"ive analysis yields a $p$-value of $9.88\times10^{-9}$. Both the total number of constraints and the solution time increase substantially with $\kappa$, whereas the no-decay critical value $\eta^*(1)$ and the critical ratio decrease as $\kappa$ increases. 
\begin{table}[H]
\centering
\begin{tabular}{lrrr}
\toprule
 & $\kappa=2$ & $\kappa=3$ & $\kappa=4$ \\
\midrule
Total constraints & 4,492,815  & 13,463,401 & 32,793,310\\
\quad Control--control & 3,756,935 & 11,319,064  & 27,498,440 \\
\quad Treated--treated & 731,186  & 2,139,643  & 5,290,176 \\
\quad Matched-set size & 4,694 & 4,694 & 4,694 \\
Solver time (seconds) & 128.09 & 731.50  & 11,816.25 \\
\midrule
Randomization inference with design-based bounds $p$-value & 0.487 & 0.625 & 0.713 \\
No-decay critical $\eta^*(1)$ & 0.113 & 0.056 & 0.031 \\
Relevant matched-set pairs & 10,891 & 25,503 & 45,443 \\
Critical ratio (\%) & 12.9 & 6.0 & 3.1 \\
\bottomrule
\end{tabular}
\caption{Summary of computational and inferential results under different thresholds.}
\label{tab:fhs-threshold-inference}
\end{table}

\newpage
\putbib
\clearpage

\end{appendix}
\end{bibunit}

\end{document}